\documentclass[prd,nofootinbib,twocolumn]{revtex4}
\usepackage{amsfonts}
\usepackage{amsmath}
\usepackage{graphicx}
\usepackage{amssymb}

\begin{document}

\title{Spinning compact binary dynamics and chameleon orbits}
\author{L\'{a}szl\'{o} \'{A}rp\'{a}d Gergely, Zolt\'{a}n Keresztes}
\affiliation{Departments of Theoretical and Experimental Physics, University of Szeged,
Hungary{\small {\qquad } }}

\begin{abstract}
We analyse the conservative evolution of spinning compact binaries to second
post-Newtonian (2PN) order accuracy, with leading order spin-orbit,
spin-spin and mass quadrupole-monopole contributions included. As a main
result we derive a closed system of first order differential equations in a
compact form, for a set of dimensionless variables encompassing both orbital
elements and spin angles. These evolutions are constrained by conservation
laws holding at 2PN\ order. As required by the generic theory of constrained
dynamical systems we perform a consistency check and prove that the
constraints are preserved by the evolution. We apply the formalism to show
the existence of chameleon orbits, whose local, orbital parameters evolve
from elliptic (in the Newtonian sense) near pericenter, towards hyperbolic
at large distances. This behavior is consistent with the picture that
General Relativity predicts stronger gravity at short distances than
Newtonian theory does.
\end{abstract}

\date{\today }
\maketitle

\section{Introduction}

The orbital dynamics of compact objects (black holes or neutron stars)
provides one of the best testbeds of any gravitational theory \cite{Yunes}.
Such systems are characterised by violently changing mass quadrupole moment,
hence leading to emission of gravitational radiation. Gravitational waves
represent ripples in space-time curvature. The way to separate them from the
background curvature is to look for the fast changing component of the
curvature, in the high-frequency / geometric optics approximation \cite%
{Isaacson}. These perturbations of the background geometry then travel away
with the speed of light.

In certain cases they can be described in terms of a post-Newtonian (PN)
approach, excellently reviewed in Refs. \cite{Blanchet,FlanaganHughes}. Such
an approach is restricted to i) a weak-field regime, where gravity is weak
compared to its strength at a black hole horizon and the distance$\ r$ is
large compared to the horizon radius, $\varepsilon =Gm/c^{2}r\ll 1$ (here $%
\varepsilon $ is the PN parameter, $G$ the gravitational constant, $c$ the
speed of light in vacuum, $m$ and$\ r$ are the total mass and separation),
and ii) slow motions compared to the speed of light, \thinspace $v\ll c$.
Due to the virial theorem, $\varepsilon \approx v^{2}/c^{2}$. Gravitational
wave characteristics can be reliably computed in the framework of the PN
formalism throughout the inspiral (lasting from tiny values of $\varepsilon $
to an upper value of the order of $0.1$), after which a merger regime
follows (requiring a numerical analysis of the full nonlinear Einstein
equations) and a ringdown of the finally merged object (which can be
described in terms of black hole perturbation theory \cite{ringdown}).
Alternatively the full inspiral-merger-ringdown process is well encompassed
by the effective one body approach \cite{EOB}, where the waveforms are
calibrated by accurate numerical relativity simulations \cite{numerical}.

Efforts for the direct detection of gravitational waves emitted by compact
binaries by the ground-based interferometric detectors LIGO \cite{LIGO},
Kagra \cite{Kagra}, Virgo \cite{Virgo} are under way and a detection is
expected in a few years, based on the best estimated coalescence rates \cite%
{rates}. Gravitational waves emitted by supermassive black hole binary
coalescence could be observed by Pulsar Timing Arrays \cite{PulsarTiming} or
with LISA \cite{LISA} (launch date 2032).

During the inspiral, up to 2PN orders the dynamics is conservative. A
classical tests of general relativity, the perihelion shift of planetary
orbits is an 1PN effect. In the strong gravity regimes even without
including the dissipative effects arising at 2.5PN orders and beyond \cite%
{nonspinning}, the orbits could be extremely different from Keplerian ones.
As an example we mention zoom-whirl orbits, arising in geodesic calculations 
\cite{Kennefick,zoomwhirlGeod} and numerical relativity simulations \cite%
{zoomwhirlNum,Capture}. Corresponding contributions, that capture some of
these features, arise at large PN parameters \cite{zoomwhirlPN}. In the
conservative dynamics general relativistic effects contribute at 1PN and 2PN
orders, but the spins of the components also couple with the orbital angular
momentum and with themselves, leading to spin-orbit (SO) and spin-spin (SS)
effects. The mass quadrupole of the compact objects also modifies the binary
orbits with quadrupole-monopole couplings (QM). It is usual to trace back
the quadrupole moment to rotation, in case of which the QM contributions to
dynamics could be expressed in terms of the spin. These corrections to the
dynamics have been discussed extensively in Refs. \cite%
{DD,BOC,KWW,Kidder,Kepler,Inspiral1,Inspiral2}, while the backreaction of
gravitational radiation in Refs. \cite%
{PMPeters,Kidder,SO,SS,spinspin,Poisson,quadrup,QM}, and implications on
galactic jets in Refs. \cite{spinspin,XRG}.

In this paper we study the conservative dynamics of a spinning compact
binary system. We rewrite the full set of conservative evolution equations,
first derived in Refs. \cite{Inspiral1}-\cite{Inspiral2}, in terms of a set
of dimensionless variables evolving in a dimensionless time, in a form
suitable to monitor \textit{both bounded and unbounded orbits }(as defined
by their osculating orbital elements in a local, Newtonian sense).

In Section \ref{Var} we introduce the dimensionless variables closely tied
to the leading order dynamics. These include a) dynamical quantities, which
up to the 2PN conservative evolution are constants, b) angular variables
defined in reference systems selected by the orbital motion and total
angular momentum, finally c) spin angles. Then in Section \ref{Perturb} we
introduce the perturbing force of the Keplerian evolution, encompassing
corrections from general relativity in the form of the 1PN and 2PN
contributions and spin related corrections, namely the leading order SO, SS
and QM couplings. The contributions to the precessional evolutions are also
enlisted.

The full 2PN\ conservative dynamics is presented in Section \ref{2PNdynamics}
in the form of a generalisation of the Lagrange planetary equations. First
the evolution of 5 dimensionless orbital elements and of 4 spin angles is
given in terms of the true anomaly. This is complemented by the evolution of
the true anomaly in terms of the dimensionless time. As a result we obtain a
closed system of first order differential equations. They are involved,
nevertheless they exhibit a simple structure. Suitable notations made this
structure transparent.

The dimensionless variables however do not evolve unconstrained. At the 2PN
accuracy of the conservative motion there are constants of motion,
expressible in terms of these variables. We give these constraints in terms
of the dimensionless dynamical variables in Section \ref{constraints}.

As for any constrained dynamical system, consistency checks need to be
performed. This implies to take the time derivative of the constraints and
to investigate their role from a dynamical point of view. In principle such
constraints could lead to A) new equations of motion, B) new constraints, or
C) identities. Section \ref{consistency} is devoted to this involved
analysis, with some of the computational details shifted to Appendix \ref%
{EnJ2PN}. We prove that the dynamical equations given in Section \ref%
{2PNdynamics} are exhaustive in describing the binary and spin evolution, as
the time derivatives of the constraints lead to identities. We fulfil the
task by performing a series of consistency checks of the system of
differential equations at each PN and spin order.

As an application of the derived formalism in Section \ref%
{eccentrichyperbolic} we analyse the possibility of having orbits which
change from hyperbolic to elliptic and vice-versa, in terms of the
eccentricity of the osculating ellipse, thus in a Newtonian sense. The
existence of such evolutions are to be expected, as General Relativity
predicts stronger gravity at short distances than the Newtonian theory, as
well known from the study of the stellar equilibrium. Indeed, we find orbits
dubbed \textit{chameleon}, which appear elliptic (locally, in a Newtonian
sense) at short range, but transform into hyperbolic (in the same sense) at
larger distances. Finally Section \ref{CoRe} contains the concluding remarks.

Throughout the paper an overhat denotes the direction of the respective
vector.

\section{Dimensionless variables\label{Var}}

\subsection{Dynamical characteristics of axially symmetric compact objects 
\label{massspinqm}}

Compact binary components with axial symmetry are characterised by their
mass $m_{i}$, their proper rotation encompassed in their dimensionless spin $%
\chi _{i}$ and their quadrupole coefficient $w_{i}$.

The mass of neutron stars is typically of $1.4$ solar masses. Black holes on
the other hand could have masses extending from a few solar masses (for
stellar mass black holes), up to $10^{10}$ solar masses (for the largest
mass supermassive black holes).\ We will frequently employ the total and
reduced masses $m=m_{1}+m_{2}$ and $\mu =m_{1}m_{2}/m$, the mass ratio $\nu
=m_{2}/m_{1}\in \left[ 0,1\right] $ and its symmetrical counterpart $\eta
=\mu /m\in \left[ 0,0.25\right] $.

Kerr black holes in extreme rotation provide the upper bound of the
dimensionless spin parameter, which for general relativistic black holes is
constrained as $\chi _{i}\in \left[ 0,1\right] $. Faster rotation would
destroy the horizon, rendering them into naked singularities. In order to
estimate the range of $\chi _{i}$ for neutron stars, we proceed as follows.
From the expression of the spin magnitudes $S_{i}=\left( G/c\right)
m_{i}^{2}\chi _{i}$ we rewrite the dimensionless spin as a ratio of two
dimensionless parameters: $\chi _{i}=(S_{i}/m_{i}R_{i}c)/(Gm_{i}/c^{2}R_{i})$%
. For a neutron star of $1.4$ solar masses and radius of $10$ km the
denominator is $(Gm_{i}/c^{2}R_{i})\approx 0.2$. Approximating the neutron
star to leading order by a rigid sphere, the numerator becomes $%
(S_{i}/m_{i}R_{i}c)\approx $ $(2/5)(R_{i}\Omega _{i}/c)$, hence $\chi
_{i}\approx 2(R_{i}\Omega _{i}/c)$. Unless the surface rotational velocity $%
R_{i}\Omega _{i}$ of the neutron star is higher than half of the speed of
light (typical observed rotational velocities are much smaller), for neutron
stars $\chi _{i}\in $ $\left[ 0,1\right] $ also holds.\footnote{%
We note that this is not the case for ordinary stars, where $%
(Gm_{i}/c^{2}R_{i})$ has a much smaller value due to their non-compactness.
Hence in the dynamics spin-orbit coupling terms containing $\chi _{i}$ could
dominate over general relativistic corrections, while spin-spin terms with $%
\chi _{1}\chi _{2}$ or quadrupole-monopole terms with $\chi _{i}^{2}$ could
become even larger.}

When the quadrupole moment arises from rotation rather than asymmetric mass
distribution, $w_{i}=1$ holds for general relativistic black holes \cite%
{Thorne 1980}, while for neutron stars $w_{i}\in \left[ 4,8\right] $,
depending on their equation of state \cite{Poisson}, \cite{Larakkers Poisson
1997}.

\subsection{Keplerian dynamical constants \label{const_evol}}

The Newtonian expressions of the energy, orbital angular momentum and
Laplace-Runge-Lenz vector of a binary system are%
\begin{eqnarray}
E_{N} &\equiv &\frac{\mu v^{2}}{2}-\frac{Gm\mu }{r},  \label{ENdef} \\
\mathbf{L}_{\mathbf{N}} &\equiv &\mu \mathbf{r}\times \mathbf{v~},
\label{LNdef} \\
\mathbf{A}_{\mathbf{N}} &\equiv &\mathbf{v}\times \mathbf{L}_{\mathbf{N}%
}-Gm\mu \mathbf{\hat{r}~.}  \label{ANdef}
\end{eqnarray}%
They obey the constraints 
\begin{eqnarray}
\mathbf{L}_{\mathbf{N}}\cdot \mathbf{A}_{\mathbf{N}} &=&0~,  \notag \\
\left( Gm\mu \right) ^{2}+\frac{2E_{N}L_{N}^{2}}{\mu } &=&A_{N}^{2}~,
\label{constr}
\end{eqnarray}%
and are constants to Keplerian order. As it is well-known, the Keplerian
orbit is a conic section characterised by these constants.

\subsection{Osculating orbit}

When general relativistic corrections in the weak field and slow motion
approximation are taken into account as PN and 2PN corrections; also by
including the modifications to the dynamics due to leading SO, SS and QM
couplings, the orbit ceases to be a conic section in the strict sense,
nevertheless it can be well approximated by a conic section locally. This
local approximant is the osculating orbit, defined as the Keplerian orbit
with the same orbital state vectors (position and velocity) as for the orbit
realised in the presence of the perturbations. It is easy to see then that
the dynamical constants of the osculating orbit are $E_{N},~\mathbf{L}_{%
\mathbf{N}},~\mathbf{A}_{\mathbf{N}}$, restricted by the constraints (\ref%
{constr}).

The perturbed orbit can be envisaged as a sequence of conic sections, the
orbital elements of which slowly evolve. One can therefore characterise the
osculating orbit (the instantaneous Keplerian orbit, the orbital parameters
of which evolve in time) by the above introduced 5 independent and
time-evolving variables. The additional information encoded in the orbital
state vectors is $\dot{r}$.

\subsection{Dimensionless orbital elements and spin variables}

The 5 independent dynamical variables are equivalent to a similar number of
orbital elements. To show this, first we define two independent variables
characterising the shape of the osculating orbit, which are both
dimensionless and equally apply for bounded or unbounded orbits. These are a
dimensionless version of the Newtonian orbital angular momentum and the
orbital eccentricity, defined as: 
\begin{eqnarray}
\mathfrak{l}_{r} &=&\frac{cL_{N}}{Gm\mu }~,  \label{lr} \\
e_{r} &=&\frac{A_{N}}{Gm\mu }~.  \label{er}
\end{eqnarray}%
In these variables the Newtonian expression of the energy reads (see the
second constraint (\ref{constr})):%
\begin{equation}
E_{N}=\mu c^{2}\frac{e_{r}^{2}-1}{2\mathfrak{l}_{r}^{2}}~,  \label{EN}
\end{equation}%
which is manifestly negative for circular ($e_{r}=0$) or elliptical orbits ($%
0<e_{r}<1)$, zero for parabolic orbits ($e_{r}=1$) and positive for
hyperbolic orbits ($e_{r}>1$). Note that $\mathfrak{l}_{r}$ is related to
the semi-latus rectum $p_{N}=L_{N}^{2}/Gm\mu ^{2}$ of the conic orbit as $%
\mathfrak{l}_{r}=c\left( p_{N}/Gm\right) ^{1/2}$ and to the periastron $%
r_{\min }=p_{N}/\left( 1+e_{r}\right) $ as 
\begin{equation}
\mathfrak{l}_{r}=\left( \frac{c^{2}r_{\min }}{Gm}\right) ^{1/2}\left(
1+e_{r}\right) ^{1/2}~.  \label{lrrmin}
\end{equation}%
Note that $r_{\min }=0$ and $\mathfrak{l}_{r}=0$ are equivalent, thus a
collision course is possible only for vanishing orbital angular momentum.
For bounded orbits we can also introduce the semimajor axis $%
a_{r}=p_{N}/\left( 1-e_{r}^{2}\right) =\left( Gm/c^{2}\right) \mathfrak{l}%
_{r}^{2}/\left( 1-e_{r}^{2}\right) $.\footnote{%
Note that it is possible to define in a similar way a semi-latus rectum $%
p_{Kerr}$ and radial eccentricity $e_{Kerr}$ in terms of the conserved
energy and $z$-component of the orbital angular momentum of Kerr orbits, as
introduced for bounded orbits in Ref. \cite{Kennefick}.}

For the three Euler angles, defining the orientation of the plane of motion
and the orientation of the orbit in this plane we chose the following: the
inclination $\alpha =\arccos \left( \mathbf{\hat{J}\cdot \hat{L}}_{\mathbf{N}%
}\right) $ of the plane of orbit with respect to the reference plane (which
is chosen perpendicular to the total angular momentum $\mathbf{J}$), the
longitude of the ascending node $-\phi _{n}$ (measured from an arbitrary
axis $\mathbf{\hat{x}}$ lying in the reference plane to the ascending node $%
\mathbf{\hat{l}}$, defined by the intersection of the reference plane with
the plane of motion) and the argument of the periastron $\psi _{p}$
(measured from the ascending node to the periastron in the plane of motion,
see Fig 2. of Ref. \cite{Inspiral1}). These three Euler angles together with
($\mathfrak{l}_{r},e_{r}$) are equivalent with the set of dynamical
variables ($E_{N}$, $\mathbf{L}_{\mathbf{N}}$, $\mathbf{A}_{\mathbf{N}}$),
as only five of the latter are independent due to the constraints (\ref%
{constr}).

Note that the definition of the above angles is meaningful only when the
ascending node and the periastron can be defined, thus alternative
definitions of the angles are required in the cases a) of evolutions which
are either nonspinning or the spins are aligned to the orbital angular
momentum, when the ascending node cannot be defined in the above sense, and
b) for circular orbits, when there is no periastron. We regard these however
as configurations of measure zero in the generic parameter space, which need
special attention. The formalism developed in this paper is well suited for
precessing configurations and noncircular orbits.

The spin polar and azimuthal angles are $\kappa _{i}=\arccos \left( \mathbf{%
\hat{S}}_{\mathbf{i}}\cdot \mathbf{\hat{L}}_{\mathbf{N}}\right) $ and $\psi
_{i}$ (when measured from the ascending node), or $\zeta _{i}=\psi _{i}-\psi
_{p}~$(when measured from the periastron). In this paper we employ the
latter, as this will simplify the notations.

Finally we mention the last angular variable necessary for the description
of the compact binary dynamics. The position of the reduced mass particle is
characterised by its azimuthal angle, this is $\chi _{p}$ when measured from
the periastron, or $\psi _{p}+\chi _{p}$ when measured from the ascending
node. The other quantity, which defines its instantaneous position is the
distance $r$ measured from the focal point, where the potential generated by
the (fixed) total mass $m$ is centered. Its relation with the already
introduced quantities will be discussed next.

\subsection{Parametrization of the radial evolution\label{param}}

The true anomaly parametrization $r\left( \chi _{p}\right) $ of the
osculating orbit is the same as for the Keplerian motion 
\begin{equation}
r=\frac{L_{N}^{2}}{\mu \left( Gm\mu +A_{N}\cos \chi _{p}\right) }~,
\label{truer}
\end{equation}%
with the important difference that the dynamical quantities $L_{N}$ and $%
A_{N}$ evolve with the osculating orbit. The parametrization obeys%
\begin{eqnarray}
\dot{r} &=&\frac{A_{N}}{L_{N}}\sin \chi _{p}~,  \label{truerdot} \\
v^{2} &=&\frac{\left( Gm\mu \right) ^{2}+A_{N}^{2}+2Gm\mu A_{N}\cos \chi _{p}%
}{L_{N}^{2}}~.  \label{truev2}
\end{eqnarray}%
In terms of osculating orbital elements introduced above, these relations
read%
\begin{eqnarray}
r &=&\frac{Gm}{c^{2}}\frac{\mathfrak{l}_{r}^{2}}{1+e_{r}\cos \chi _{p}}~,
\label{truerosc} \\
\dot{r} &=&c\frac{e_{r}}{\mathfrak{l}_{r}}\sin \chi _{p}~,
\label{truerdotosc} \\
v^{2} &=&c^{2}\frac{1+e_{r}^{2}+2e_{r}\cos \chi _{p}}{\mathfrak{l}_{r}^{2}}~.
\label{truev2osc}
\end{eqnarray}

\subsection{Dimensionless constants of motion}

At the level of accuracy discussed in this paper (with PN, SO, 2PN, SS, QM
contributions to the dynamics included) there are several constants of the
motion:

(a) the magnitudes $S_{i}=\left( G/c\right) m\mu \nu ^{2i-3}\chi _{i}$ of
the spins $\mathbf{S}_{\mathbf{i}}\ $(as the spins undergo a purely
precessional evolution \cite{BOC}),

(b) the total energy $E=E_{N}+E_{PN}+E_{SO}+E_{2PN}+E_{SS}+E_{QM}$ of the
system, with the various contributions given in Refs. \cite{KWW}, \cite%
{Kidder}, \cite{quadrup},

(c) the total angular momentum $\mathbf{J=L}_{\mathbf{N}}\mathbf{+\mathbf{L}%
_{P\mathbf{N}}\mathbf{\mathbf{+}L}_{\mathbf{SO}}\mathbf{+\mathbf{L}_{2P%
\mathbf{N}}}+S}_{\mathbf{1}}\mathbf{+S}_{\mathbf{2}}$, with contributions
enlisted in Ref. \cite{Kidder}; however for the contribution $\mathbf{%
\mathbf{L}_{\mathbf{SO}}}$ we adopt the expression given in Ref. \cite%
{Inspiral2}, which holds true in the Newton-Wigner-Price \cite{NWP} spin
supplementary condition (SSC), employed in this paper.

We introduce the dimensionless versions for the total energy and angular
momentum magnitude as%
\begin{eqnarray}
\mathfrak{E} &=&\frac{E}{\mu c^{2}}~,  \label{Edimless} \\
\mathfrak{J} &=&\frac{cJ}{Gm\mu }~.  \label{Jmagdimless}
\end{eqnarray}%
Note that unlike other quantities employed in this section (characteristic
to the local approximant of the real orbit, e.g. to the osculating orbit),
the total energy and angular momentum (also their dimensionless versions)
characterise the real orbit.

It is also possible to define the periastron distance $r_{\min }^{\mathfrak{J%
}}$ and eccentricity $e_{r}^{\mathfrak{J}}$ of the fictious Keplerian motion
with energy $\mathfrak{E}$ and orbital angular momentum $\mathfrak{J}$
through the relations%
\begin{eqnarray}
\mathfrak{J}^{2} &=&\left( \frac{c^{2}r_{\min }^{\mathfrak{J}}}{Gm}\right)
\left( 1+e_{r}^{\mathfrak{J}}\right) ~, \\
A &=&Gm\mu e_{r}^{\mathfrak{J}}~,
\end{eqnarray}%
where $A$ is the Laplace-Runge-Lenz vector of the Keplerian motion with
energy $\mathfrak{E}$ and orbital angular momentum $\mathfrak{J}$, defined
in the usual way as 
\begin{equation}
A=Gm\mu \left( 1+2\mathfrak{EJ}^{2}\right) ^{1/2}~.
\end{equation}%
These relations lead to the expressions of the orbital elements of the
fictious Keplerian motion in terms of $\mathfrak{E}$ and $\mathfrak{J}$ as 
\begin{eqnarray}
e_{r}^{\mathfrak{J}} &=&\left( 1+2\mathfrak{EJ}^{2}\right) ^{1/2}~,  \notag
\\
\frac{c^{2}r_{\min }^{\mathfrak{J}}}{Gm} &=&\frac{\mathfrak{J}^{2}}{1+\left(
1+2\mathfrak{EJ}^{2}\right) ^{1/2}}~.
\end{eqnarray}%
To 2PN accuracy both $r_{\min }^{\mathfrak{J}}$ and $e_{r}^{\mathfrak{J}}$
are constants.\footnote{%
Note that similar definitions can be also introduced as 
\begin{eqnarray}
e_{r}^{\mathfrak{L}} &=&\left( 1+2\mathfrak{EL}^{2}\right) ^{1/2}~,  \notag
\\
\frac{c^{2}r_{\min }^{\mathfrak{L}}}{Gm} &=&\frac{\mathfrak{L}^{2}}{1+\left(
1+2\mathfrak{EL}^{2}\right) ^{1/2}}~,
\end{eqnarray}%
where $\mathfrak{L}=cL/(Gm\mu )$, with $L$ the magnitude of the total
orbital angular momentum. As $L$ is not a constant when SS and QM
contributions to the dynamics are present, $r_{\min }^{\mathfrak{L}}$ and $%
e_{r}^{\mathfrak{L}}$ vary on the orbit, therefore they are not particularly
useful. Alternatively, an \textit{orbital average} $\bar{L}$ of the
magnitude of orbital angular momentum was introduced and employed in Refs. 
\cite{spinspin}, \cite{quadrup} and \cite{mdipole} together with the
corresponding orbital elements in Ref. \cite{Kepler}, but only for closed
orbits.}

With this we have all ingredients to obtain the dynamics of a spinning,
precessing compact binary on noncircular orbit at 2PN order accuracy.

\section{Relativistic and spin induced perturbations\label{Perturb}}

We will characterise the deviation from Keplerian evolution in terms of a
generic perturbing force, which receives contributions from 1PN relativistic
corrections (given in terms of relative coordinates in Ref. \cite{DD}), the
2PN\ relativistic, the leading order SO (in the Newton-Wigner-Pryce SSC) and
SS corrections (all given in Ref. \cite{Kidder}) and quadrupolar
contributions (given in Ref. \cite{quadrup}). The spins, with the exception
of the aligned case also induce precessions of the spins and of the orbital
plane.

We define the dimensionless versions of the perturbing force $\Delta \mathbf{%
a}$ acting on unit mass and of the spin angular frequencies $\mathbf{\Omega }%
_{\mathbf{i}}$ as follows 
\begin{eqnarray}
\mathfrak{a} &=&\frac{Gm}{c^{4}}\Delta \mathbf{a~,}  \notag \\
\mathfrak{\omega }_{\mathbf{i}} &=&\frac{Gm}{c^{3}}\Omega _{\mathbf{i}}~.
\end{eqnarray}%
The components of $\Delta \mathbf{a}$\ and $\mathbf{\Omega }_{\mathbf{i}}$
expressed in the system $\mathbf{f}_{(\mathbf{i})}=(\mathbf{\hat{A}}_{%
\mathbf{N}},\ \mathbf{\hat{Q}}_{\mathbf{N}}\equiv \mathbf{\hat{L}}_{\mathbf{N%
}}\times \mathbf{\hat{A}}_{\mathbf{N}},~\mathbf{\hat{L}}_{\mathbf{N}})$ were
given in Appendix B of Ref. \cite{Inspiral2} in terms of the variables $%
\left( r,\dot{r},v\right) $. Starting from those, by employing the
parametrization (\ref{truerosc})-(\ref{truev2osc}) and by rewriting all
quantities in terms of the dimensionless variables introduced in the
previous section, also by switching to a description in terms of the spin
azimuthal angles $\zeta _{i}$ (rather than $\psi _{i}$), finally organising
the expressions such that they can be written in a compact form, which
emphasizes their true anomaly dependence, we explicitly give the projections
of the perturbing accelerations and spin angular frequencies below.

\subsection{Perturbing force}

The dimensionless version of the component of the perturbing force along the
periastron line is:

\begin{eqnarray}
\mathfrak{a}\cdot \mathbf{\hat{A}}_{\mathbf{N}} &=&\mathfrak{a}_{1}^{PN}+%
\mathfrak{a}_{1}^{2PN}+\mathfrak{a}_{1}^{SO}+\mathfrak{a}_{1}^{SS}+\mathfrak{%
a}_{1}^{QM}~, \\
\mathfrak{a}_{1}^{PN} &=&\frac{\left( 1+e_{r}\cos \chi _{p}\right) ^{2}}{%
\mathfrak{l}_{r}^{6}}\sum_{k=0}^{3}\mathrm{c}_{1\left( k\right) }^{PN}\cos
^{k}\chi _{p}~,  \notag \\
\mathfrak{a}_{1}^{2PN} &=&\frac{\left( 1+e_{r}\cos \chi _{p}\right) ^{2}}{%
\mathfrak{l}_{r}^{8}}\sum_{k=0}^{5}\mathrm{c}_{1\left( k\right) }^{2PN}\cos
^{k}\chi _{p}~,  \notag \\
\mathfrak{a}_{1}^{SO} &=&\frac{{\eta }\left( 1+e_{r}\cos \chi _{p}\right)
^{3}\left( e_{r}+\cos \chi _{p}\right) }{2\mathfrak{l}_{r}^{7}}  \notag \\
&&\times \sum_{k=1}^{2}\left( 4\nu ^{2k-3}+3\right) \chi _{k}\cos \kappa
_{k}~,  \notag \\
\mathfrak{a}_{1}^{SS} &=&\frac{3\eta \left( 1+e_{r}\cos \chi _{p}\right) ^{4}%
}{\mathfrak{l}_{r}^{8}}\chi _{1}\chi _{2}\Bigl\{-\cos \kappa _{1}\cos \kappa
_{2}  \notag \\
&&\times \cos \chi _{p}+\frac{1}{4}\sin \kappa _{1}\sin \kappa _{2}\left[
2\cos \zeta _{\left( -\right) }\cos \chi _{p}\right.  \notag \\
&&\left. +\cos \left( \chi _{p}-\zeta _{\left( +\right) }\right) +5\cos
\left( 3\chi _{p}-\zeta _{\left( +\right) }\right) \right] \Bigr\}~,  \notag
\\
\mathfrak{a}_{1}^{QM} &=&-\frac{3\eta \left( 1+e_{r}\cos \chi _{p}\right)
^{4}}{2\mathfrak{l}_{r}^{8}}\sum_{k=1}^{2}w_{k}\nu ^{2k-3}\chi _{k}^{2}%
\Bigl\{\cos \chi _{p}  \notag \\
&&-\frac{1}{2}\sin ^{2}\kappa _{k}\allowbreak \left[ \cos \zeta _{k}+5\cos
\left( 2\chi _{p}-\zeta _{k}\right) \right]  \notag \\
&&\times \cos \left( \chi _{p}-\zeta _{k}\right) \Bigr\}~,  \notag
\end{eqnarray}%
where $\zeta _{\left( \pm \right) }=\zeta _{2}\pm \zeta _{1}$ and the
coefficients $\mathrm{c}_{1\left( k\right) }^{PN}$ and $\mathrm{c}_{1\left(
k\right) }^{2PN}$ are given as 
\begin{eqnarray}
\mathrm{c}_{1\left( 0\right) }^{PN} &=&-2\left( 2-\eta \right) e_{r}~, 
\notag \\
\mathrm{c}_{1\left( 1\right) }^{PN} &=&\text{\textrm{C}}_{1}-\allowbreak
\left( 1+\frac{3\eta }{2}\right) e_{r}^{2}~,  \notag \\
\mathrm{c}_{1\left( 2\right) }^{PN} &=&6\left( 1-\eta \right) e_{r}~,  \notag
\\
\mathrm{c}_{1\left( 3\right) }^{PN} &=&-\allowbreak \frac{3\eta }{2}%
e_{r}^{2}~,  \label{c1PN}
\end{eqnarray}%
and 
\begin{eqnarray}
\mathrm{c}_{1\left( 0\right) }^{2PN} &=&\left[ 2+13\eta +2\eta ^{2}-\eta
\left( 3-\eta \right) e_{r}^{2}\right] e_{r}~,  \notag \\
\mathrm{c}_{1\left( 1\right) }^{2PN} &=&\text{\textrm{C}}_{2}+\left( 4+\frac{%
71\eta }{2}+2\eta ^{2}\right) e_{r}^{2}-\frac{\eta }{8}\left( 3-29\eta
\right) e_{r}^{4}~,  \notag \\
\mathrm{c}_{1\left( 2\right) }^{2PN} &=&\left[ \text{\textrm{C}}_{3}+\left(
2+\allowbreak 27\eta \right) e_{r}^{2}\right] e_{r}~,  \notag \\
\mathrm{c}_{1\left( 3\right) }^{2PN} &=&\left[ \text{\textrm{C}}_{4}-\frac{%
3\eta }{4}\left( 1+7\eta \right) e_{r}^{2}\right] \allowbreak e_{r}^{2}~, 
\notag \\
\mathrm{c}_{1\left( 4\right) }^{2PN} &=&\left( -2-\frac{59\eta }{2}+13\eta
^{2}\right) e_{r}^{3}~,  \notag \\
\mathrm{c}_{1\left( 5\right) }^{2PN} &=&-\frac{15\eta }{8}\left( 1-3\eta
\right) e_{r}^{4}~,  \label{c12PN}
\end{eqnarray}%
with the shorthand notations%
\begin{eqnarray}
\text{\textrm{C}}_{1} &=&3-\eta ~,  \notag \\
\text{\textrm{C}}_{2} &=&-9-\frac{73\eta }{4}\allowbreak +2\eta ^{2}~, 
\notag \\
\text{\textrm{C}}_{3} &=&-20-49\eta +8\eta ^{2}~,  \notag \\
\text{\textrm{C}}_{4} &=&-13-\frac{223\eta }{4}+16\eta ^{2}~.
\end{eqnarray}

The dimensionless version of the perturbing force component in the plane of
motion, but perpendicular to the periastron line is

\begin{eqnarray}
\mathfrak{a}\cdot \mathbf{\hat{Q}}_{\mathbf{N}} &=&\mathfrak{a}_{2}^{PN}+%
\mathfrak{a}_{2}^{2PN}+\mathfrak{a}_{2}^{SO}+\mathfrak{a}_{2}^{SS}+\mathfrak{%
a}_{2}^{QM}~, \\
\mathfrak{a}_{2}^{PN} &=&\frac{\left( 1+e_{r}\cos \chi _{p}\right) ^{2}\sin
\chi _{p}}{\mathfrak{l}_{r}^{6}}\sum_{k=0}^{2}\mathrm{c}_{2\left( k\right)
}^{PN}\cos ^{k}\chi _{p}~,  \notag \\
\mathfrak{a}_{2}^{2PN} &=&\frac{\left( 1+e_{r}\cos \chi _{p}\right) ^{2}\sin
\chi _{p}}{\mathfrak{l}_{r}^{8}}\sum_{k=0}^{4}\mathrm{c}_{2\left( k\right)
}^{2PN}\cos ^{k}\chi _{p}~,  \notag \\
\mathfrak{a}_{2}^{SO} &=&\frac{\eta \left( 1+e_{r}\cos \chi _{p}\right)
^{3}\sin \chi _{p}}{2\mathfrak{l}_{r}^{7}}  \notag \\
&&\times \sum_{k=1}^{2}\left( 4\nu ^{2k-3}+3\right) \chi _{k}\cos \kappa
_{k}~,  \notag
\end{eqnarray}%
\begin{eqnarray}
\mathfrak{a}_{2}^{SS} &=&\frac{3\eta \left( 1+e_{r}\cos \chi _{p}\right) ^{4}%
}{\mathfrak{l}_{r}^{8}}\chi _{1}\chi _{2}\Bigl\{-\cos \kappa _{1}\cos \kappa
_{2}  \notag \\
&&\times \sin \chi _{p}+\frac{1}{4}\sin \kappa _{1}\sin \kappa _{2}\left[
2\cos \zeta _{\left( -\right) }\sin \chi _{p}\right.  \notag \\
&&\left. -\sin \left( \chi _{p}-\zeta _{\left( +\right) }\right) +5\sin
\left( 3\chi _{p}-\zeta _{\left( +\right) }\right) \right] \Bigr\}~,  \notag
\\
\mathfrak{a}_{2}^{QM} &=&-\frac{3\eta \left( 1+e_{r}\cos \chi _{p}\right)
^{4}}{2\mathfrak{l}_{r}^{8}}\sum_{k=1}^{2}w_{k}\nu ^{2k-3}\chi _{k}^{2}%
\Bigl\{\sin \chi _{p}  \notag \\
&&-\frac{1}{2}\sin ^{2}\kappa _{k}\left[ \sin \zeta _{k}+5\sin \left( 2\chi
_{p}-\zeta _{k}\right) \right]  \notag \\
&&\times \cos \left( \chi _{p}-\zeta _{k}\right) \Bigr\}~,  \notag
\end{eqnarray}%
where the coefficients $\mathrm{c}_{2\left( k\right) }^{PN}$ and $\mathrm{c}%
_{2\left( k\right) }^{2PN}$ are given as%
\begin{eqnarray}
\mathrm{c}_{2\left( 0\right) }^{PN} &=&\text{\textrm{C}}_{1}+\left( 3-\frac{%
7\eta }{2}\right) e_{r}^{2}~,  \notag \\
\mathrm{c}_{2\left( 1\right) }^{PN} &=&\mathrm{c}_{1\left( 2\right) }^{PN}~,
\notag \\
\mathrm{c}_{2\left( 2\right) }^{PN} &=&\mathrm{c}_{1\left( 3\right) }^{PN}~,
\label{c2PN}
\end{eqnarray}%
and%
\begin{eqnarray}
\mathrm{c}_{2\left( 0\right) }^{2PN} &=&\text{\textrm{C}}_{2}+17\eta
e_{r}^{2}+\frac{21\eta }{8}\left( 1+\allowbreak \eta \right) e_{r}^{4}~, 
\notag \\
\mathrm{c}_{2\left( 1\right) }^{2PN} &=&\left[ \text{\textrm{C}}_{3}+\left(
26+3\eta \right) \eta e_{r}^{2}\right] e_{r}~,  \notag \\
\mathrm{c}_{2\left( 2\right) }^{2PN} &=&\left[ \text{\textrm{C}}_{4}+\frac{%
3\eta }{4}\left( 5-3\eta \right) e_{r}^{2}\right] e_{r}^{2}~,  \notag \\
\mathrm{c}_{2\left( 3\right) }^{2PN} &=&\mathrm{c}_{1\left( 4\right)
}^{2PN}~,  \notag \\
\mathrm{c}_{2\left( 4\right) }^{2PN} &=&\mathrm{c}_{1\left( 5\right)
}^{2PN}~.  \label{c22PN}
\end{eqnarray}

The dimensionless version of the perturbing force component perpendicular to
the plane of motion has only spin induced contributions%
\begin{eqnarray}
\mathfrak{a}\cdot \mathbf{\hat{L}}_{\mathbf{N}} &=&\mathfrak{a}_{3}^{SO}+%
\mathfrak{a}_{3}^{SS}+\mathfrak{a}_{3}^{QM}~, \\
\mathfrak{a}_{3}^{SO} &=&\frac{{\eta }\left( 1+e_{r}\cos \chi _{p}\right)
^{3}}{4\mathfrak{l}_{r}^{7}}\sum_{k=1}^{2}\left( 4\nu ^{2k-3}+3\right) \chi
_{k}\sin \kappa _{k}  \notag \\
&&\times \left[ e_{r}\allowbreak \cos \zeta _{k}+4\cos \left( \chi
_{p}-\zeta _{k}\right) \right.  \notag \\
&&\left. +3e_{r}\cos \left( 2\chi _{p}-\zeta _{k}\right) \right] ~,  \notag
\\
\mathfrak{a}_{3}^{SS} &=&-\frac{3\eta \left( 1+e_{r}\cos \chi _{p}\right)
^{4}}{\mathfrak{l}_{r}^{8}}\chi _{1}\chi _{2}\left[ \cos \kappa _{1}\sin
\kappa _{2}\right.  \notag \\
&&\left. \times \cos \left( \chi _{p}-\zeta _{2}\right) +\cos \kappa
_{2}\sin \kappa _{1}\cos \left( \chi _{p}-\zeta _{1}\right) \right] ~, 
\notag \\
\mathfrak{a}_{3}^{QM} &=&-\frac{3\eta \left( 1+e_{r}\cos \chi _{p}\right)
^{4}}{2\mathfrak{l}_{r}^{8}}  \notag \\
&&\times \sum_{k=1}^{2}w_{k}\nu ^{2k-3}\chi _{k}^{2}\sin 2\kappa _{k}\cos
\left( \chi _{p}-\zeta _{k}\right) ~.  \notag
\end{eqnarray}

An important remark we make here is that the PN order of various terms can
be evaluated from the relative powers of $\mathfrak{l}_{r}$ in the
respective terms, $\mathfrak{l}_{r}^{-1}$ counting for 0.5PN orders. As $%
\mathfrak{l}_{r}$ is much larger than unity, it is also much larger than the
dimensionless spins $\chi _{i}$.

\subsection{The precessional angular velocities}

The precessions arise due to the SO, SS and QM contributions to the
dynamics. The components of the dimensionless angular velocity are:%
\begin{eqnarray}
\mathfrak{\omega }_{\mathbf{i}}\cdot \mathbf{\hat{A}}_{\mathbf{N}} &=&\frac{%
\eta \left( 1+e_{r}\cos \chi _{p}\right) ^{3}}{2\mathfrak{l}_{r}^{6}}\left\{
\nu ^{2j-3}\chi _{j}\sin \kappa _{j}\right.  \notag \\
&&\times \left[ 3\cos \left( 2\chi _{p}-\zeta _{j}\right) +\cos \zeta _{j}%
\right] +3w_{i}\chi _{i}  \notag \\
&&\left. \times \sin \kappa _{i}\left[ \cos \left( 2\chi _{p}-\zeta
_{i}\right) +\cos \zeta _{i}\right] \right\} ~, \\
\mathfrak{\omega }_{\mathbf{i}}\cdot \mathbf{\hat{Q}}_{\mathbf{N}} &=&\frac{%
\eta \left( 1+e_{r}\cos \chi _{p}\right) ^{3}}{2\mathfrak{l}_{r}^{6}}\left\{
\nu ^{2j-3}\chi _{j}\sin \kappa _{j}\right.  \notag \\
&&\times \left[ 3\sin \left( 2\chi _{p}-\zeta _{j}\right) +\sin \zeta _{j}%
\right] +3w_{i}\chi _{i}  \notag \\
&&\left. \times \sin \kappa _{i}\left[ \sin \left( 2\chi _{p}-\zeta
_{i}\right) +\sin \zeta _{i}\right] \right\} \mathbf{~,} \\
\mathfrak{\omega }_{\mathbf{i}}\cdot \mathbf{\hat{L}}_{\mathbf{N}} &=&\frac{%
\eta \left( 1+e_{r}\cos \chi _{p}\right) ^{3}}{2\mathfrak{l}_{r}^{5}}\left(
4+3\nu ^{3-2i}\right)  \notag \\
&&-\frac{\eta \left( 1+e_{r}\cos \chi _{p}\right) ^{3}}{2\mathfrak{l}_{r}^{6}%
}\nu ^{2j-3}\chi _{j}\cos \kappa _{j}~,  \label{Om3}
\end{eqnarray}%
with $j\neq i$. The first term in $\mathfrak{\omega }_{\mathbf{i}}\cdot 
\mathbf{\hat{L}}_{\mathbf{N}}$ is due to the SO interaction. The terms
containing $w_{i}$ are due to the QM\ interaction, while the other terms
containining $\chi _{i}$ due to the SS interaction.

Note that all terms in the equations above carry the same power of the PN
parameter $\mathfrak{l}_{r}^{-2}$. Whether any of the terms dominate,
depends on the mass ratio.

\section{2PN conservative dynamics\label{2PNdynamics}}

In Ref. \cite{Inspiral2} the evolutions of the independent variables were
derived as a system of first order coupled ordinary differential equations.
We rewrite below these evolutions explicitly in terms of the dimensionless
variables of this paper. We will also switch from $\psi _{i}$ to $\zeta _{i}$%
, and switch to a dimensionless time variable, defined as%
\begin{equation}
\mathfrak{t}=\frac{c^{3}}{Gm}t~.
\end{equation}%
We will denote by an overdot the derivative with respect to $\mathfrak{t}$
(as opposed to Ref. \cite{Inspiral2}, where an overdot was the derivative
with respect to $t$): 
\begin{equation}
\frac{d}{d\mathfrak{t}}=\frac{Gm}{c^{3}}\frac{d}{dt}~.
\end{equation}%
With all these changes in the notation and in the choice of independent
variables the equations simplify considerably.

For the osculating orbital elements we obtain the coupled system of the
evolutions:%
\begin{eqnarray}
\mathfrak{\dot{l}}_{r} &=&\frac{\mathfrak{l}_{r}^{2}}{1+e_{r}\cos \chi _{p}}%
\bigl[-\left( \mathfrak{a}\cdot \mathbf{\hat{A}}_{\mathbf{N}}\right) \sin
\chi _{p}  \notag \\
&&+\left( \mathfrak{a}\cdot \mathbf{\hat{Q}}_{\mathbf{N}}\right) \cos \chi
_{p}\bigr]\mathbf{\ ,}  \label{lrdot}
\end{eqnarray}%
\begin{eqnarray}
\dot{e}_{r} &=&\frac{\mathfrak{l}_{r}}{1+e_{r}\cos \chi _{p}}\bigl[-\left( 
\mathfrak{a}\cdot \mathbf{\hat{A}}_{\mathbf{N}}\right) \left( e_{r}+\cos
\chi _{p}\right) \sin \chi _{p}  \notag \\
&&+\left( \mathfrak{a}\cdot \mathbf{\hat{Q}}_{\mathbf{N}}\right) \left(
1+2e_{r}\cos \chi _{p}+\cos ^{2}\chi _{p}\right) \bigr]\mathbf{\ ,}
\label{erdot}
\end{eqnarray}%
\begin{eqnarray}
\dot{\psi}_{p} &=&-\frac{\mathfrak{l}_{r}}{\left( 1+e_{r}\cos \chi
_{p}\right) }\Bigl[\left( \mathfrak{a}\cdot \mathbf{\hat{L}}_{\mathbf{N}%
}\right) \frac{\sin \left( \psi _{p}+\chi _{p}\right) }{\tan \alpha }  \notag
\\
&&+\left( \mathfrak{a}\cdot \mathbf{\hat{A}}_{\mathbf{N}}\right) \frac{%
\left( 1+e_{r}\cos \chi _{p}+\sin ^{2}\chi _{p}\right) \!}{e_{r}}  \notag \\
&&-\left( \mathfrak{a}\cdot \mathbf{\hat{Q}}_{\mathbf{N}}\right) \frac{%
\!\sin \chi _{p}\cos \chi _{p}}{e_{r}}\Bigr]~,  \label{psipdot}
\end{eqnarray}%
\begin{equation}
\dot{\alpha}=\mathfrak{l}_{r}\left( \mathfrak{a}\cdot \mathbf{\hat{L}}_{%
\mathbf{N}}\right) \frac{\cos \left( \psi _{p}+\chi _{p}\right) }{%
1+e_{r}\cos \chi _{p}}~,  \label{alphadot}
\end{equation}%
\begin{equation}
\dot{\phi}_{n}=-\mathfrak{l}_{r}\left( \mathfrak{a}\cdot \mathbf{\hat{L}}_{%
\mathbf{N}}\right) \frac{\sin \left( \psi _{p}+\chi _{p}\right) }{\left(
1+e_{r}\cos \chi _{p}\right) \sin \alpha }~.  \label{phindot}
\end{equation}%
The spin angles evolve as:%
\begin{eqnarray}
~\dot{\kappa}_{i} &=&-\left( \mathfrak{\omega }_{\mathbf{i}}\cdot \mathbf{%
\hat{A}}_{\mathbf{N}}\right) \sin \zeta _{i}+\left( \mathfrak{\omega }_{%
\mathbf{i}}\cdot \mathbf{\hat{Q}}_{\mathbf{N}}\right) \cos \zeta _{i}  \notag
\\
&&-\mathfrak{l}_{r}\left( \mathfrak{a}\cdot \mathbf{\hat{L}}_{\mathbf{N}%
}\right) \frac{\sin \left( \chi _{p}-\zeta _{i}\right) }{1+e_{r}\cos \chi
_{p}}~,  \label{kappaidot}
\end{eqnarray}%
\begin{eqnarray}
\dot{\zeta}_{i} &=&-\left[ \left( \mathfrak{\omega }_{\mathbf{i}}\cdot 
\mathbf{\hat{A}}_{\mathbf{N}}\right) \cos \zeta _{i}+\left( \mathfrak{\omega 
}_{\mathbf{i}}\cdot \mathbf{\hat{Q}}_{\mathbf{N}}\right) \sin \zeta _{i}%
\right] \cot \kappa _{i}  \notag \\
&&+\left( \mathfrak{\omega }_{\mathbf{i}}\cdot \mathbf{\hat{L}}_{\mathbf{N}%
}\right) +\frac{\mathfrak{l}_{r}}{\left( 1+e_{r}\cos \chi _{p}\right) } 
\notag \\
&&\times \Bigl[\left( \mathfrak{a}\cdot \mathbf{\hat{L}}_{\mathbf{N}}\right) 
\frac{\cos \left( \chi _{p}-\zeta _{i}\right) }{\tan \kappa _{i}}  \notag \\
&&+\left( \mathfrak{a}\cdot \mathbf{\hat{A}}_{\mathbf{N}}\right) \frac{%
\left( 1+e_{r}\cos \chi _{p}+\sin ^{2}\chi _{p}\right) \!}{e_{r}}  \notag \\
&&-\left( \mathfrak{a}\cdot \mathbf{\hat{Q}}_{\mathbf{N}}\right) \frac{%
\!\sin \chi _{p}\cos \chi _{p}}{e_{r}}\Bigr]\mathbf{~,}  \label{zetaidot}
\end{eqnarray}%
while the true anomaly%
\begin{eqnarray}
\dot{\chi}_{p} &=&\frac{\left( 1+e_{r}\cos \chi _{p}\right) ^{2}}{\mathfrak{l%
}_{r}^{3}}+\frac{\mathfrak{l}_{r}}{e_{r}\left( 1+e_{r}\cos \chi _{p}\right) }
\notag \\
&&\times \Bigl[\left( \mathfrak{a}\cdot \mathbf{\hat{A}}_{\mathbf{N}}\right)
\left( 1+e_{r}\cos \chi _{p}+\sin ^{2}\chi _{p}\right)  \notag \\
&&-\left( \mathfrak{a}\cdot \mathbf{\hat{Q}}_{\mathbf{N}}\right) \sin \chi
_{p}\cos \chi _{p}\Bigr]~.  \label{chipdot}
\end{eqnarray}%
This latter equation allows to replace (dimensionless) time-derivatives with
derivatives with respect to $\chi _{p}$ in all previous evolution equations.

Although not independent from the previous ones, for completeness we also
give the evolutions of the auxiliary spin azimuthal angles $\psi _{i}$:%
\begin{eqnarray}
\dot{\psi}_{i} &=&-\left[ \left( \mathfrak{\omega }_{\mathbf{i}}\cdot 
\mathbf{\hat{A}}_{\mathbf{N}}\right) \cos \zeta _{i}+\left( \mathfrak{\omega 
}_{\mathbf{i}}\cdot \mathbf{\hat{Q}}_{\mathbf{N}}\right) \sin \zeta _{i}%
\right] \cot \kappa _{i}  \notag \\
&&+\left( \mathfrak{\omega }_{\mathbf{i}}\cdot \mathbf{\hat{L}}_{\mathbf{N}%
}\right) -\frac{\mathfrak{l}_{r}}{1+e_{r}\cos \chi _{p}}\left( \mathfrak{a}%
\cdot \mathbf{\hat{L}}_{\mathbf{N}}\right)  \notag \\
&&\times \left[ \frac{\sin \left( \chi _{p}+\psi _{p}\right) }{\tan \alpha }-%
\frac{\cos \left( \chi _{p}-\zeta _{i}\right) }{\tan \kappa _{i}}\right] 
\mathbf{~,}  \label{psiidot}
\end{eqnarray}%
and of the auxiliary angle $\gamma $ span by the spin vectors:%
\begin{eqnarray}
\sin \gamma ~\dot{\gamma} &=&\left( \mathfrak{\omega }_{\left( \mathbf{-}%
\right) }\cdot \mathbf{\hat{A}}_{\mathbf{N}}\right) \left[ -\cos \kappa
_{1}\sin \kappa _{2}\sin \zeta _{2}\right.  \notag \\
&&\left. +\sin \kappa _{1}\cos \kappa _{2}\sin \zeta _{1}\right] +\left( 
\mathfrak{\omega }_{\left( \mathbf{-}\right) }\cdot \mathbf{\hat{Q}}_{%
\mathbf{N}}\right)  \notag \\
&&\times \left[ \cos \kappa _{1}\sin \kappa _{2}\cos \zeta _{2}-\sin \kappa
_{1}\cos \kappa _{2}\cos \zeta _{1}\right]  \notag \\
&&+\left( \mathfrak{\omega }_{\left( \mathbf{-}\right) }\cdot \mathbf{\hat{L}%
}_{\mathbf{N}}\right) ~\sin \kappa _{1}\sin \kappa _{2}\sin \zeta _{\left(
-\right) }~.  \label{gammadot}
\end{eqnarray}%
Here we denoted $\mathfrak{\omega }_{\left( \mathbf{-}\right) }=\mathfrak{%
\omega }_{\mathbf{2}}-\mathfrak{\omega }_{\mathbf{1}}$.

These evolution equations in terms of dimensionless variables stand as the
main result of the paper.

\section{Constraints on the variables\label{constraints}}

At 2PN order accuracy, with the leading order SO, SS and QM contributions
included, the total energy and total angular momentum are conserved. These
primary constraints can be expressed in terms of the dimensionless dynamical
variables for which we derived evolution equations. Therefore in this
section we derive these constraints.

\subsection{Total energy}

Starting from the expression of the total energy, with the PN and 2PN
contributions explicitly given in \cite{Kidder}, SO (in the
Newton-Wigner-Price SSC) in \cite{Kepler}, SS in \cite{spinspin} and QM in 
\cite{quadrup}, rewritten in the notations of the present paper as\footnote{$%
G$ and $c$ were reintroduced in all 1PN and 2PN terms on dimensional
grounds. In the SS and QM terms $\chi $ of Refs. \cite{spinspin}\ and \cite%
{quadrup} was replaced with $\chi _{p}$, as to leading order they agree.
Also $\psi _{0}~$is denoted in this paper as $\psi _{p}$ and $\delta
_{i}=2\left( \psi _{0}-\psi _{i}\right) $ as $-2\zeta _{i}$.}%
\begin{eqnarray}
E_{PN} &=&\mu c^{2}\Bigl\{\frac{3}{8}\left( 1-3\eta \right) \frac{v^{4}}{%
c^{4}}+\frac{1}{2}\left( 3+\eta \right) \frac{Gm}{c^{2}r}\frac{v^{2}}{c^{2}}
\notag \\
&&+\frac{\eta }{2}\frac{Gm}{c^{2}r}\frac{\dot{r}^{2}}{c^{2}}+\frac{1}{2}%
\left( \frac{Gm}{c^{2}r}\right) ^{2}\Bigr\}~,  \notag
\end{eqnarray}%
\begin{eqnarray}
E_{2PN} &=&\mu c^{2}\Bigl\{\frac{5}{16}\left( 1-7\eta +13\eta ^{2}\right) 
\frac{v^{6}}{c^{6}}  \notag \\
&&-\frac{3}{8}\eta \left( 1-3\eta \right) \frac{Gm}{c^{2}r}\frac{\dot{r}^{4}%
}{c^{4}}  \notag \\
&&+\frac{1}{8}\left( 21-23\eta -27\eta ^{2}\right) \frac{Gm}{c^{2}r}\frac{%
v^{4}}{c^{4}}  \notag \\
&&+\frac{1}{8}\left( 14-55\eta +4\eta ^{2}\right) \left( \frac{Gm}{c^{2}r}%
\right) ^{2}\frac{v^{2}}{c^{2}}  \notag \\
&&+\frac{\eta }{4}\left( 1-15\eta \right) \frac{Gm}{c^{2}r}\frac{v^{2}}{c^{2}%
}\frac{\dot{r}^{2}}{c^{2}}-\frac{1}{4}\left( 2+15\eta \right) \left( \frac{Gm%
}{c^{2}r}\right) ^{3}  \notag \\
&&+\frac{1}{8}\left( 4+69\eta +12\eta ^{2}\right) \left( \frac{Gm}{c^{2}r}%
\right) ^{2}\frac{\dot{r}^{2}}{c^{2}}\Bigr\}~,  \notag
\end{eqnarray}%
\begin{equation*}
E_{SO}=0~,
\end{equation*}%
\begin{eqnarray}
E_{SS} &=&-\frac{G^{3}m^{4}\eta ^{2}}{2c^{4}r^{3}}\chi _{1}\chi _{2}\bigl\{%
3\cos \kappa _{1}\cos \kappa _{2}-\cos \gamma  \notag \\
&&-3\sin \kappa _{1}\sin \kappa _{2}\cos \left( 2\chi _{p}-\zeta _{\left(
+\right) }\right) \bigr\}~,  \notag
\end{eqnarray}%
\begin{eqnarray}
E_{QM} &=&-\frac{G^{3}m^{4}\eta ^{2}}{2c^{4}r^{3}}\!\!\sum_{i=1}^{2}w_{i}%
\chi _{i}^{2}\nu ^{2i-3}  \notag \\
&&\times \left[ 1-\!3\!\sin ^{2}\!\kappa _{i}\cos ^{2}\!\left( \!\chi
\!_{p}-\zeta _{i}\right) \right] ~,
\end{eqnarray}%
with $\gamma $ related to the other variables by the spherical cosine
identity 
\begin{equation*}
\cos \gamma =\cos \kappa _{1}\cos \kappa _{2}+\sin \kappa _{1}\sin \kappa
_{2}\cos \zeta _{\left( -\right) }~,
\end{equation*}%
we find that the osculating orbital elements, spin variables and true
anomaly obey the constraint%
\begin{equation}
\mathfrak{E}=\mathfrak{E}_{N}+\mathfrak{E}_{PN}+\mathfrak{E}_{2PN}+\mathfrak{%
E}_{SS}+\mathfrak{E}_{QM}~,  \label{constren}
\end{equation}%
with the contributions%
\begin{equation}
\mathfrak{E}_{N}=\frac{e_{r}^{2}-1}{2\mathfrak{l}_{r}^{2}}~,
\label{endimlessNdet}
\end{equation}%
\begin{eqnarray}
\mathfrak{E}_{PN} &=&\frac{1}{8\mathfrak{l}_{r}^{4}}\sum_{k=0}^{3}\mathrm{q}%
_{k}e_{r}^{k}\cos ^{k}\chi _{p}~,  \label{endimlessPNdet} \\
\mathrm{q}_{0} &=&19-5\eta +2\left( 9-5\eta \right) e_{r}^{2}+3\left(
1-3\eta \right) e_{r}^{4}~,  \notag \\
\mathrm{q}_{1} &=&4\left[ 14-6\eta +\left( 6-7\allowbreak \eta \right)
e_{r}^{2}\right] ~,  \notag \\
\mathrm{q}_{2} &=&8\left( 5-4\eta \right) ~,  \notag \\
\mathrm{q}_{3} &=&-4\eta ~,  \notag
\end{eqnarray}%
\begin{eqnarray}
\mathfrak{E}_{2PN} &=&\frac{1}{16\mathfrak{l}_{r}^{6}}\sum_{k=0}^{5}\mathrm{s%
}_{k}e_{r}^{k}\cos ^{k}\chi _{p}~,  \label{endimless2PNdet} \\
\mathrm{s}_{0} &=&67-\allowbreak 251\eta +19\eta ^{2}+\left( 135-165\eta
+59\allowbreak \eta ^{2}\right) e_{r}^{2}  \notag \\
&&+3\left( 19-\allowbreak 51\eta +33\eta ^{2}\right) e_{r}^{4}  \notag \\
&&+5\left( 1-7\eta +13\eta ^{2}\right) e_{r}^{6}~,  \notag \\
\mathrm{s}_{1} &=&\allowbreak 2\left[ 2\left( 82-265\eta +38\eta ^{2}\right)
\right.  \notag \\
&&+\allowbreak 2\left( 96-157\eta +85\eta ^{2}\right) e_{r}^{2}  \notag \\
&&\left. +3\left( 12-43\eta \allowbreak +\allowbreak 49\eta ^{2}\right)
e_{r}^{4}\right] ~,  \notag \\
\mathrm{s}_{2} &=&4\left[ 126-\allowbreak 415\eta +106\eta ^{2}\right. 
\notag \\
&&\left. +\allowbreak \left( 66-140\eta +125\eta ^{2}\right) e_{r}^{2}\right]
~,  \notag \\
\mathrm{s}_{3} &=&4\left[ 60-258\eta +113\eta ^{2}+2\eta \left( 1+3\eta
\right) e_{r}^{2}\right] ~,  \notag \\
\mathrm{s}_{4} &=&-2\left( 4+76\eta -57\eta ^{2}\right) ~,  \notag \\
\mathrm{s}_{5} &=&\frac{16}{5e_{r}^{4}}\allowbreak \mathrm{c}_{2\left(
4\right) }^{2PN}~.  \notag
\end{eqnarray}%
\begin{eqnarray}
\mathfrak{E}_{SS} &=&-\frac{\eta \left( 1+e_{r}\cos \chi _{p}\right) ^{3}}{2%
\mathfrak{l}_{r}^{6}}\chi _{1}\chi _{2}  \notag \\
&&\times \Bigl\{2\cos \kappa _{1}\cos \kappa _{2}-\sin \kappa _{1}\sin
\kappa _{2}  \notag \\
&&\times \left[ \cos \zeta _{\left( -\right) }+3\cos \left( 2\chi _{p}-\zeta
_{\left( +\right) }\right) \right] \Bigr\}~,  \label{endimlessSSdet}
\end{eqnarray}%
and%
\begin{eqnarray}
&&\mathfrak{E}_{QM}=-\frac{\eta \left( 1+e_{r}\cos \chi _{p}\right) ^{3}}{2%
\mathfrak{l}_{r}^{6}}\sum_{i=1}^{2}w_{i}\chi _{i}^{2}\nu ^{2i-3}  \notag \\
&&\times \left[ 1-\!3\!\sin ^{2}\!\kappa _{i}\cos ^{2}\!\left( \!\chi
\!_{p}-\zeta _{i}\right) \right] ~.  \label{endimlessQMdet}
\end{eqnarray}

\subsection{Total angular momentum}

The projections along the basis vectors $\left( \mathbf{\hat{l}},~\mathbf{%
\hat{m}\equiv \hat{L}}_{\mathbf{N}}\times \mathbf{\hat{l}},~\mathbf{\hat{L}}%
_{\mathbf{N}}\right) $ of the expression of the total angular momentum give
constraint relations. In the Newton-Wigner-Price SSC these were given as
Eqs. (B26)-(B28) of \cite{Inspiral2}. We rewrite these relations in terms of
the dimensionless variables employed in this paper, also employ
trigonometric identities to give them in the most simple form containing the
spin azimuthal angles $\zeta _{i}$ (rather than $\psi _{i}$). We obtain%
\begin{align}
0& =\sum_{i=1}^{2}\chi _{i}\sin \kappa _{i}\Bigl[\nu ^{2i-3}\cos \left(
\zeta _{i}+\psi _{p}\right)  \notag \\
& -\frac{\eta \left( 4\nu ^{2i-3}+3\right) \left( 1+e_{r}\cos \chi
_{p}\right) }{2\mathfrak{l}_{r}^{2}}  \notag \\
& \times \sin \left( \chi _{p}+\psi _{p}\right) \sin \left( \chi _{p}-\zeta
_{i}\right) \Bigr]\ ,  \label{constr1}
\end{align}%
\begin{align}
\mathfrak{J}\sin \alpha & =\sum_{i=1}^{2}\chi _{i}\sin \kappa _{i}\Bigl[\nu
^{2i-3}\sin \left( \zeta _{i}+\psi _{p}\right)  \notag \\
& +\frac{\eta \left( 4\nu ^{2i-3}+3\right) \left( 1+e_{r}\cos \chi
_{p}\right) }{2\mathfrak{l}_{r}^{2}}  \notag \\
& \times \cos \left( \chi _{p}+\psi _{p}\right) \sin \left( \chi _{p}-\zeta
_{i}\right) \Bigr]\ ,  \label{constr2}
\end{align}%
\begin{align}
\mathfrak{J}\cos \alpha & =\mathfrak{l}_{r}\left( 1+\epsilon _{PN}+\epsilon
_{2PN}\right) +\sum_{i=1}^{2}\chi _{i}\cos \kappa _{i}  \notag \\
& \times \Bigl[\nu ^{2i-3}-\frac{\eta \left( 4\nu ^{2i-3}+3\right) \left(
1+e_{r}\cos \chi _{p}\right) }{2\mathfrak{l}_{r}^{2}}\Bigr]\ ,
\label{constr3}
\end{align}%
with%
\begin{equation}
\epsilon _{PN}=\frac{7-\eta +\left( 1-3\eta \right) e_{r}^{2}+4\left( 2-\eta
\right) e_{r}\cos \chi _{p}}{2\mathfrak{l}_{r}^{2}}~,  \label{epPNdet}
\end{equation}%
and%
\begin{eqnarray}
\epsilon _{2PN} &=&\frac{1}{8\mathfrak{l}_{r}^{4}}\sum_{k=0}^{3}\mathrm{p}%
_{k}e_{r}^{k}\cos ^{k}\chi _{p}~,  \label{ep2PNdet} \\
\mathrm{p}_{0} &=&59-143\eta +11\eta ^{2}+2\left( 17\allowbreak -45\eta
+11\eta ^{2}\right) e_{r}^{2}  \notag \\
&&+3\left( 1-7\eta +13\eta ^{2}\right) e_{r}^{4}~,  \notag \\
\mathrm{p}_{1} &=&4\left( 38-\allowbreak 92\eta +16\eta ^{2}+\left(
10-\allowbreak 33\eta +25\eta ^{2}\right) e_{r}^{2}\right) ~,  \notag \\
\mathrm{p}_{2} &=&2\left( \allowbreak 48-119\eta +56\eta ^{2}\right) ~, 
\notag \\
\mathrm{p}_{3} &=&4\eta \left( 2+5\eta \right) ~.  \notag
\end{eqnarray}%
For aligned configurations the constraints (\ref{constr1})-(\ref{constr2})
become identities.

We note that all three total angular momentum constraints have a leading
order and an $\mathfrak{l}_{r}^{-2}$ contribution. It is instructive to
discuss the leading order contributions to Eqs. (\ref{constr1})-(\ref%
{constr3}):%
\begin{eqnarray}
0 &=&\sum_{i=1}^{2}\chi _{i}\sin \kappa _{i}\nu ^{2i-3}\cos \psi _{i}+%
\mathcal{O}\left( \mathfrak{l}_{r}^{-2}\right) ~, \\
\mathfrak{J}\sin \alpha &=&\sum_{i=1}^{2}\chi _{i}\sin \kappa _{i}\nu
^{2i-3}\sin \psi _{i}+\mathcal{O}\left( \mathfrak{l}_{r}^{-2}\right) \ , \\
\mathfrak{J}\cos \alpha &=&\mathfrak{l}_{r}+\sum_{i=1}^{2}\chi _{i}\cos
\kappa _{i}\nu ^{2i-3}+\mathcal{O}\left( \mathfrak{l}_{r}^{-2}\right) \ ,
\end{eqnarray}%
while the ratio of the last two becomes%
\begin{equation}
\tan \alpha =\frac{\sum_{i=1}^{2}\frac{\chi _{i}}{\mathfrak{l}_{r}}\sin
\kappa _{i}\nu ^{2i-3}\sin \psi _{i}}{1+\sum_{i=1}^{2}\frac{\chi _{i}}{%
\mathfrak{l}_{r}}\cos \kappa _{i}\nu ^{2i-3}}+\mathcal{O}\left( \mathfrak{l}%
_{r}^{-2}\right) \ .
\end{equation}%
As $\chi _{i}/\mathfrak{l}_{r}=\mathcal{O}\left( \varepsilon ^{1/2}\right) $%
, for comparable masses $\tan \alpha =\mathcal{O}\left( \varepsilon
^{1/2}\right) \sin \kappa _{i}$, hence $\alpha $ is of 0.5PN order. By
contrast, for small mass ratios $\tan \alpha =\tan \kappa _{1}/\left[ 1+%
\mathcal{O}\left( \varepsilon ^{-1/2}\right) \nu \right] $ which is
approximated as $\alpha \approx \kappa _{1}$ (not necessarily small) when $%
\mathcal{O}\left( \varepsilon ^{-1/2}\right) \nu \ll 1$.

Another useful formula holding to leading order, which will be explored
later on in the paper is: 
\begin{equation}
\mathfrak{l}_{r}=\sum_{i=1}^{2}\nu ^{2i-3}\chi _{i}\left[ \sin \kappa
_{i}\sin \left( \zeta _{i}+\psi _{p}\right) \cot \alpha -\cos \kappa _{i}%
\right] +\mathcal{O}\left( \mathfrak{l}_{r}^{-2}\right) \ .  \label{lrconstr}
\end{equation}%
For comparable masses $\sin \kappa _{i}\cot \alpha =\mathcal{O}\left( 
\mathfrak{l}_{r}\right) $, hence it is large, while $\cos \kappa _{i}$ is of
order unity. By contrast, for small mass ratios, where $\alpha \approx
\kappa _{1}$, the two terms are of comparable unit order, nevertheless the
prefactor $\nu ^{-1}\chi _{1}$ is large, of order $\mathfrak{l}_{r}$.

The equations (\ref{constren}), (\ref{constr1})-(\ref{constr3}) are primary
constraints for the 2PN accurate binary dynamics, the consistency of which
with the dynamical equations has to be analysed. This will be done in the
next section.

\section{Consistency\label{consistency}}

According to the general theory of dynamical systems with constraints, the
derivatives of the constraints could lead to either new dynamical equations,
new constraints, or be identically satisfied. In case of new constraints
arising by this procedure, the check of the consistency conditions should be
repeated. Therefore in the present section we discuss these consistency
conditions.

We will verify the consistency of the above lengthy system of evolution and
constraint equations by taking the time derivatives of the four dynamical
constraints (\ref{constren}), (\ref{constr1})-(\ref{constr3}) derived in the
previous section and inserting in them the evolution of the orbital elements
and spin angles given in Section \ref{2PNdynamics}. We will do this order by
order, starting with the Keplerian order, then proceeding with the
relativistic 1PN and 2PN contributions, finally discussing the leading order
consistency for the SO, SS and QM contributions. The calculations will be
somewhat simplified by taking into account that only $\chi _{p}$ has a
Newtonian order evolution; the orbital elements and spin angles being
conserved at this order.

\subsection{Time derivative of the total angular momentum constraint}

For calculating the time derivative of the total angular momentum constraint
one has to remember that the basis $\left( \mathbf{\hat{l}},~\mathbf{\hat{m}%
\equiv \hat{L}}_{\mathbf{N}}\times \mathbf{\hat{l}},~\mathbf{\hat{L}}_{%
\mathbf{N}}\right) $ employed for the decomposition (\ref{constr1})-(\ref%
{constr3}) itself changes, being a precessing basis. Hence for the
consistency condition we need to prove 
\begin{equation}
0=\frac{d}{d\mathfrak{t}}\left( \mathfrak{J}_{\mathbf{\hat{l}}}\mathbf{\hat{l%
}}+\mathfrak{J}_{\mathbf{\hat{m}}}\mathbf{\hat{m}}+\mathfrak{J}_{\mathbf{%
\hat{L}}_{\mathbf{N}}}\mathbf{\hat{L}}_{\mathbf{N}}\right) ~.
\label{JSOconsistency}
\end{equation}

For the evolutions of the basis vectors we start from the precession
relations (12)-(13) given in Ref. \cite{Inspiral2} for the basis $\mathbf{f}%
_{(\mathbf{i})}$, and rewrite them in terms of the dimensionless variables
as:

\begin{equation}
\mathbf{\dot{f}}_{(\mathbf{i})}=\mathbf{\Omega }_{A}\times \mathbf{f}_{(%
\mathbf{i})}~,  \label{KAdirevol}
\end{equation}%
with the angular velocity vector (redefined by a factor of $dt/d\mathfrak{t}$
as compared to Ref. \cite{Inspiral2}) 
\begin{eqnarray}
\mathbf{\Omega }_{A} &=&\frac{\mathfrak{l}_{r}}{1+e_{r}\cos \chi _{p}}%
\Biggl\{\left( \mathfrak{a}\cdot \mathbf{\hat{L}}_{\mathbf{N}}\right) \left(
\cos \chi _{p}\mathbf{\hat{A}}_{\mathbf{N}}+\sin \chi _{p}\mathbf{\hat{Q}}_{%
\mathbf{N}}\right)  \notag \\
&&-\frac{1}{e_{r}}\Bigl[\left( \mathfrak{a}\cdot \mathbf{\hat{A}}_{\mathbf{N}%
}\right) \left( 2+e_{r}\cos \chi _{p}-\cos ^{2}\chi _{p}\right)  \notag \\
&&\!-\!\left( \mathfrak{a}\cdot \mathbf{\hat{Q}}_{\mathbf{N}}\right) \sin
\chi _{p}\cos \chi _{p}\Bigr]\mathbf{\hat{L}}_{\mathbf{N}}\Biggr\}~.
\label{OmegaAVec1}
\end{eqnarray}%
Next we take into account that the basis$\left( \mathbf{\hat{l}},\mathbf{%
\hat{m}}\right) $ is transformed into $\left( \mathbf{\hat{A}}_{\mathbf{N}},%
\mathbf{\hat{Q}}_{\mathbf{N}}\right) $ by a rotation with angle $\psi _{p}$:%
\begin{eqnarray}
\mathbf{\hat{l}} &=&\cos \psi _{p}\mathbf{\hat{A}}_{\mathbf{N}}-\sin \psi
_{p}\mathbf{\hat{Q}}_{\mathbf{N}}~,  \label{l} \\
\mathbf{\hat{m}} &=&\sin \psi _{p}\mathbf{\hat{A}}_{\mathbf{N}}+\cos \psi
_{p}\mathbf{\hat{Q}}_{\mathbf{N}}~,  \label{m}
\end{eqnarray}
which leads to a precession of the basis vectors $\left( \mathbf{\hat{l},%
\hat{m},\hat{L}}_{\mathbf{N}}\right) $ with the angular frequency vector%
\begin{equation}
\mathbf{\Omega }_{L}=\mathbf{\Omega }_{A}-\dot{\psi}_{p}\mathbf{\hat{L}}_{%
\mathbf{N}}~.  \label{OmegaLVec0}
\end{equation}%
(Note, that in contrast with the expression (24) given in Ref. \cite%
{Inspiral2} here the dot refers to the derivative with respect to the
dimensionless time $\mathfrak{t}$.) The detailed form of $\mathbf{\Omega }%
_{L}$ was also given as Eq. (30) in Ref. \cite{Inspiral2}, which, after a
proper rescaling to account for the evolution in terms of $\mathfrak{t}$ and
rewritten in terms of the dimensionless variables reads%
\begin{eqnarray}
\mathbf{\Omega }_{L} &=&\frac{\mathfrak{l}_{r}\left( \mathfrak{a}\cdot 
\mathbf{\hat{L}}_{\mathbf{N}}\right) }{1+e_{r}\cos \chi _{p}}\Bigl[\cos \chi
_{p}\mathbf{\hat{A}}_{\mathbf{N}}+\sin \chi _{p}\mathbf{\hat{Q}}_{\mathbf{N}}
\notag \\
&&+\frac{\sin \left( \psi _{p}+\chi _{p}\right) }{\tan \alpha }\mathbf{\hat{L%
}}_{\mathbf{N}}\Bigr]~,  \label{OmegaLVec1}
\end{eqnarray}%
or rewritten in the basis $\left( \mathbf{\hat{l},\hat{m},\hat{L}}_{\mathbf{N%
}}\right) $ as%
\begin{eqnarray}
\mathbf{\Omega }_{L} &=&\frac{\mathfrak{l}_{r}\left( \mathfrak{a}\cdot 
\mathbf{\hat{L}}_{\mathbf{N}}\right) }{1+e_{r}\cos \chi _{p}}\Bigl[\cos
\left( \psi _{p}+\chi _{p}\right) ~\mathbf{\hat{l}}+\sin \left( \psi
_{p}+\chi _{p}\right) ~\mathbf{\hat{m}}  \notag \\
&&+\frac{\sin \left( \psi _{p}+\chi _{p}\right) }{\tan \alpha }\mathbf{\hat{L%
}}_{\mathbf{N}}\Bigr]~.  \label{OmegaLVec2}
\end{eqnarray}

Hence the consistency condition to be proven reads%
\begin{eqnarray}
0 &=&\mathfrak{\dot{J}}_{\mathbf{\hat{l}}}\mathbf{\hat{l}}+\mathfrak{\dot{J}}%
_{\mathbf{\hat{m}}}\mathbf{\hat{m}}+\mathfrak{\dot{J}}_{\mathbf{\hat{L}}_{%
\mathbf{N}}}\mathbf{\hat{L}}_{\mathbf{N}}  \notag \\
&&+\mathfrak{J}_{\mathbf{\hat{l}}}\mathbf{\Omega }_{L}\times \mathbf{\hat{l}}%
+\mathfrak{J}_{\mathbf{\hat{m}}}\mathbf{\Omega }_{L}\times \mathbf{\hat{m}}+%
\mathfrak{J}_{\mathbf{\hat{L}}_{\mathbf{N}}}\mathbf{\Omega }_{L}\times 
\mathbf{\hat{L}}_{\mathbf{N}}~.
\end{eqnarray}%
We rewrite this by inserting the components of the normalized total angular
momentum $\left( \mathfrak{J}_{\mathbf{\hat{l}}}\mathbf{,}\mathfrak{J}_{%
\mathbf{\hat{m}}}\mathbf{,}\mathfrak{J}_{\mathbf{\hat{L}}_{\mathbf{N}%
}}\right) =\left( 0,\mathfrak{J}\sin \alpha ,\mathfrak{J}\cos \alpha \right) 
$ and by exploring Eqs. (\ref{m}) and (\ref{OmegaLVec2}):%
\begin{eqnarray}
0 &=&\mathfrak{\dot{J}}_{\mathbf{\hat{l}}}\mathbf{\hat{l}}+\mathfrak{\dot{J}}%
_{\mathbf{\hat{m}}}\mathbf{\hat{m}}+\mathfrak{\dot{J}}_{\mathbf{\hat{L}}_{%
\mathbf{N}}}\mathbf{\hat{L}}_{\mathbf{N}}+\mathfrak{J}\frac{\mathfrak{l}%
_{r}\left( \mathfrak{a}\cdot \mathbf{\hat{L}}_{\mathbf{N}}\right) }{%
1+e_{r}\cos \chi _{p}}  \notag \\
&&\times \cos \left( \psi _{p}+\chi _{p}\right) \left( -\cos \alpha \ 
\mathbf{\hat{m}+}\sin \alpha \ \mathbf{\hat{L}}_{\mathbf{N}}\right) ~.
\end{eqnarray}%
Hence the desired consistency conditions are 
\begin{eqnarray}
0 &=&\mathfrak{\dot{J}}_{\mathbf{\hat{l}}}~,  \label{consJdotl} \\
0 &=&\mathfrak{\dot{J}}_{\mathbf{\hat{m}}}-\mathfrak{J}\cos \alpha \ \frac{%
\mathfrak{l}_{r}\left( \mathfrak{a}\cdot \mathbf{\hat{L}}_{\mathbf{N}%
}\right) }{1+e_{r}\cos \chi _{p}}\cos \left( \psi _{p}+\chi _{p}\right) ~,
\label{consJdotm} \\
0 &=&\mathfrak{\dot{J}}_{\mathbf{\hat{L}}_{\mathbf{N}}}+\mathfrak{J}\sin
\alpha ~\frac{\mathfrak{l}_{r}\left( \mathfrak{a}\cdot \mathbf{\hat{L}}_{%
\mathbf{N}}\right) }{1+e_{r}\cos \chi _{p}}\cos \left( \psi _{p}+\chi
_{p}\right) ~.  \label{consJdotLN}
\end{eqnarray}%
In order to prove them, for the derivatives of the normalized total angular
momentum we take the derivatives of the rhs of the constraints (\ref{constr1}%
)-(\ref{constr3}).

Note that the component $\mathfrak{\dot{J}}_{\mathbf{\hat{l}}}$ of the
normalized total angular momentum (which vanishes for nonprecessing
evolutions) is a conserved scalar for precessing evolutions. We have seen at
the end of Section \ref{constraints} that this constraint decouples into two
independent conditions, each obeyed by one of the spin directions.

Another remark is that the second terms of the rhs in Eqs. (\ref{consJdotm}%
)-(\ref{consJdotLN}) are the sign flipped versions of the derivatives of the
lhs expressions of the constraints (\ref{constr2})-(\ref{constr3}), with $%
\dot{\alpha}$ taken from Eq. (\ref{alphadot}). Hence the same consistency
conditions could be obtained by simply taking the dimensionless time
derivative of the constraints (\ref{constr2})-(\ref{constr3}).

\subsection{Keplerian evolution}

With only the leading order terms due to the vanishing of $\mathfrak{a}$ and 
$\mathfrak{\omega }_{\mathbf{i}}$, Eq. (\ref{chipdot}) reduces to 
\begin{equation}
\dot{\chi}_{p}=\frac{\left( 1+e_{r}\cos \chi _{p}\right) ^{2}}{\mathfrak{l}%
_{r}^{3}}~.
\end{equation}%
Combining this with the definition of the true anomaly, Eq. (\ref{truerosc})
we obtain Kepler's second law for the area:%
\begin{equation}
r^{2}\dot{\chi}_{p}=\frac{G^{2}m^{2}}{c^{4}}\mathfrak{l}_{r}~.
\end{equation}%
The constraint equations reduce to 
\begin{eqnarray}
\mathfrak{E} &=&\frac{e_{r}^{2}-1}{2\mathfrak{l}_{r}^{2}}~,  \notag \\
\alpha &=&0~,  \notag \\
\mathfrak{J} &=&\mathfrak{l}_{r}~.
\end{eqnarray}%
Then Eq. (\ref{alphadot}) becomes an identity, Eqs. (\ref{lrdot})-(\ref%
{erdot}), (\ref{kappaidot})-(\ref{zetaidot}) imply constant $\mathfrak{l}%
_{r} $, $e_{r}$, $\kappa _{i}$ and $\zeta _{i}$ (although at this accuracy
there are no spins, thus $\kappa _{i}$ and $\zeta _{i}$ have no
interpretation).

With $\alpha =0$, Eqs. (\ref{psipdot}) and (\ref{phindot}) become
ill-defined, unless we multiply them with $\sin \alpha $, when they give
identities, but no information on $\psi _{p}$ and $\phi _{n}$. This is
related to the ill-definedness of the node line $\mathbf{\hat{l}}$ when the
two planes coincide. Therefore some $\mathbf{\hat{l}}$ has to be chosen in
an arbitrary way to define the argument of the periastron.

This last remark also holds when only the 1PN or 1PN+2PN contributions are
included, or when the spins are perpendicular to the orbit ($\pm \mathbf{S}_{%
\mathbf{1}}\parallel \pm \mathbf{S}_{\mathbf{2}}\parallel \mathbf{L}_{%
\mathbf{N}}$) thus they do not precess. In these cases by definition $\alpha
=0$, consistent with $\mathfrak{a}\cdot \mathbf{\hat{L}}_{\mathbf{N}}=0$
(when spins are present then due to $\kappa _{i}=0$) in Eq. (\ref{alphadot}%
). For all these cases the reference plane and node line should be defined
by another vector, not aligned to $\mathbf{J}$.

\subsection{2PN level consistency, non-spinning case}

We discuss the 1PN and 2PN consistency conditions together below, by
switching off the spin.

\subsubsection{The energy condition}

The time derivative of the total energy, the constraint equation (\ref%
{constren}), without the spin and quadrupole contributions, to 2PN accuracy
gives%
\begin{equation}
0=\dot{e}_{r}\frac{e_{r}}{\mathfrak{l}_{r}^{2}}-\frac{\mathfrak{\dot{l}}_{r}%
}{\mathfrak{l}_{r}}\frac{e_{r}^{2}-1}{\mathfrak{l}_{r}^{2}}+\frac{d}{d%
\mathfrak{t}}\mathfrak{E}_{PN}+\frac{d}{d\mathfrak{t}}\mathfrak{E}_{2PN}~,
\label{consEn2PN}
\end{equation}%
with%
\begin{eqnarray}
\frac{d}{d\mathfrak{t}}\mathfrak{E}_{PN} &=&\frac{1}{8\mathfrak{l}_{r}^{4}}%
\Biggl\{-4\frac{\mathfrak{\dot{l}}_{r}}{\mathfrak{l}_{r}}\sum_{k=0}^{3}%
\mathrm{q}_{k}e_{r}^{k}\cos ^{k}\chi _{p}  \notag \\
&&+\dot{e}_{r}\sum_{k=0}^{3}\frac{d\left( \mathrm{q}_{k}e_{r}^{k}\right) }{%
de_{r}}\cos ^{k}\chi _{p}  \notag \\
&&-\dot{\chi}_{p}\sin \chi _{p}\sum_{k=0}^{3}k\mathrm{q}_{k}e_{r}^{k}\cos
^{k-1}\chi _{p}\Biggr\}~,
\end{eqnarray}%
and%
\begin{eqnarray}
\frac{d}{d\mathfrak{t}}\mathfrak{E}_{2PN} &=&\frac{1}{16\mathfrak{l}_{r}^{6}}%
\Biggl\{-6\frac{\mathfrak{\dot{l}}_{r}}{\mathfrak{l}_{r}}\sum_{k=0}^{5}%
\mathrm{s}_{k}e_{r}^{k}\cos ^{k}\chi _{p}  \notag \\
&&+\dot{e}_{r}\sum_{k=0}^{5}\frac{d\left( \mathrm{s}_{k}e_{r}^{k}\right) }{%
de_{r}}\cos ^{k}\chi _{p}  \notag \\
&&-\dot{\chi}_{p}\sin \chi _{p}\sum_{k=0}^{5}k\mathrm{s}_{k}e_{r}^{k}\cos
^{k-1}\chi _{p}\Biggl\}~,
\end{eqnarray}%
The 1PN and 2PN contributions to $\dot{e}_{r}$ and $\mathfrak{\dot{l}}_{r}/%
\mathfrak{l}_{r}$ carry factors of $\mathfrak{l}_{r}^{-5}$ and $\mathfrak{l}%
_{r}^{-7}$, respectively; while $\dot{\chi}_{p}$ has Newtonian, 1PN and 2PN
contributions, carrying factors of $\mathfrak{l}_{r}^{-3}$, $\mathfrak{l}%
_{r}^{-5}$ and $\mathfrak{l}_{r}^{-7}$, respectively. Remembering that $%
\mathfrak{l}_{r}^{-2}$ represents one relative PN order it is easy to
separate the 1PN and 2PN contributions to the consistency condition (\ref%
{consEn2PN}). These are the terms scaling with $\mathfrak{l}_{r}^{-7}$ and $%
\mathfrak{l}_{r}^{-9}$, respectively (while higher order terms should be
dropped, being beyond the accuracy of the present calculations). We find at
1PN%
\begin{eqnarray}
0 &=&\dot{e}_{r}^{PN}\frac{e_{r}}{\mathfrak{l}_{r}^{2}}-\frac{\mathfrak{\dot{%
l}}_{r}^{PN}}{\mathfrak{l}_{r}}\frac{\left( e_{r}^{2}-1\right) }{\mathfrak{l}%
_{r}^{2}}  \notag \\
&&-\frac{\dot{\chi}_{p}^{N}\sin \chi _{p}}{8\mathfrak{l}_{r}^{4}}%
\sum_{k=0}^{3}k\mathrm{q}_{k}e_{r}^{k}\cos ^{k-1}\chi _{p}~,
\label{consEn2PN1}
\end{eqnarray}%
and at 2PN%
\begin{eqnarray}
0 &=&\dot{e}_{r}^{2PN}\frac{e_{r}}{\mathfrak{l}_{r}^{2}}+\dot{e}_{r}^{PN}%
\frac{1}{8\mathfrak{l}_{r}^{4}}\sum_{k=0}^{3}\frac{d\left( \mathrm{q}%
_{k}e_{r}^{k}\right) }{de_{r}}\cos ^{k}\chi _{p}  \notag \\
&&-\frac{\mathfrak{\dot{l}}_{r}^{2PN}}{\mathfrak{l}_{r}}\frac{e_{r}^{2}-1}{%
\mathfrak{l}_{r}^{2}}-\frac{\mathfrak{\dot{l}}_{r}^{PN}}{\mathfrak{l}_{r}}%
\frac{1}{2\mathfrak{l}_{r}^{4}}\sum_{k=0}^{3}\mathrm{q}_{k}e_{r}^{k}\cos
^{k}\chi _{p}  \notag \\
&&-\frac{\sin \chi _{p}}{8\mathfrak{l}_{r}^{4}}\Biggl(\dot{\chi}%
_{p}^{PN}\sum_{k=0}^{3}k\mathrm{q}_{k}e_{r}^{k}\cos ^{k-1}\chi _{p}  \notag
\\
&&+\frac{\dot{\chi}_{p}^{N}}{2\mathfrak{l}_{r}^{2}}\sum_{k=0}^{5}k\mathrm{s}%
_{k}e_{r}^{k}\cos ^{k-1}\chi _{p}\Biggr)~.  \label{consEn2PN2}
\end{eqnarray}%
Inserting the evolutions of $e_{r}$, $\mathfrak{l}_{r}$ and $\chi _{p}$, the
1PN accurate consistency condition (\ref{consEn2PN1}) becomes%
\begin{eqnarray}
0 &=&\mathfrak{a}_{1}^{PN}\sin \chi _{p}-\mathfrak{a}_{2}^{PN}\left(
e_{r}+\cos \chi _{p}\right)  \notag \\
&&+\frac{\left( 1+e_{r}\cos \chi _{p}\right) ^{2}\sin \chi _{p}}{8\mathfrak{l%
}_{r}^{6}}\sum_{k=0}^{3}k\mathrm{q}_{k}e_{r}^{k}\cos ^{k-1}\chi _{p}~.
\end{eqnarray}%
Inserting $\mathfrak{a}_{1}^{PN}$ and $\mathfrak{a}_{2}^{PN}$ we obtain for
the coefficients of the powers $0,1,2$ and $3$ of the arbitrary $\cos \chi
_{p}$ the relations%
\begin{eqnarray}
\mathrm{q}_{1}e_{r} &=&8\left( -\mathrm{c}_{1\left( 0\right) }^{PN}+\mathrm{c%
}_{2\left( 0\right) }^{PN}e_{r}\right) ~,  \notag \\
\mathrm{q}_{2}e_{r}^{2} &=&4\left( -\mathrm{c}_{1\left( 1\right) }^{PN}+%
\mathrm{c}_{2\left( 1\right) }^{PN}e_{r}+\mathrm{c}_{2\left( 0\right)
}^{PN}\right) ~,  \notag \\
\mathrm{q}_{3}e_{r}^{3} &=&\frac{8}{3}\left( -\mathrm{c}_{1\left( 2\right)
}^{PN}+\mathrm{c}_{2\left( 2\right) }^{PN}e_{r}+\mathrm{c}_{2\left( 1\right)
}^{PN}\right) ~,  \notag \\
0 &=&\mathrm{c}_{1\left( 3\right) }^{PN}-\mathrm{c}_{2\left( 2\right)
}^{PN}~,
\end{eqnarray}%
which can easily be verified to hold with the definitions (\ref{c1PN}), (\ref%
{c2PN}) and (\ref{endimlessPNdet}) of this paper.

As expected, the 2PN part of the consistency condition (\ref{consEn2PN}),
Eq. (\ref{consEn2PN2}) gives a much more cumbersome equation:%
\begin{eqnarray}
0 &=&\left( 1+e_{r}\cos \chi _{p}\right) \left[ \mathfrak{a}_{1}^{2PN}\sin
\chi _{p}-\mathfrak{a}_{2}^{2PN}\left( e_{r}+\cos \chi _{p}\right) \right] 
\notag \\
&&+\frac{\mathfrak{a}_{1}^{PN}\sin \chi _{p}}{8\mathfrak{l}_{r}^{2}}\bigl[%
-4\sum_{k=0}^{3}\mathrm{q}_{k}e_{r}^{k}\cos ^{k}\chi _{p}  \notag \\
&&+\left( e_{r}+\cos \chi _{p}\right) \sum_{k=0}^{3}\frac{d\left( \mathrm{q}%
_{k}e_{r}^{k}\right) }{de_{r}}\cos ^{k}\chi _{p}  \notag \\
&&+\left( 1+e_{r}\cos \chi _{p}+\sin ^{2}\chi _{p}\right) \sum_{k=0}^{3}k%
\mathrm{q}_{k}\left( e_{r}\cos \chi _{p}\right) ^{k-1}\bigr]  \notag \\
&&+\frac{\mathfrak{a}_{2}^{PN}}{8\mathfrak{l}_{r}^{2}}\bigl[4\cos \chi
_{p}\sum_{k=0}^{3}\mathrm{q}_{k}e_{r}^{k}\cos ^{k}\chi _{p}  \notag \\
&&-\sin ^{2}\chi _{p}\cos \chi _{p}\sum_{k=0}^{3}k\mathrm{q}_{k}\left(
e_{r}\cos \chi _{p}\right) ^{k-1}  \notag \\
&&-\left( 1+2e_{r}\cos \chi _{p}+\cos ^{2}\chi _{p}\right) \sum_{k=0}^{3}%
\frac{d\left( \mathrm{q}_{k}e_{r}^{k}\right) }{de_{r}}\cos ^{k}\chi _{p}%
\bigr]  \notag \\
&&+\frac{\left( 1+e_{r}\cos \chi _{p}\right) ^{3}\sin \chi _{p}}{16\mathfrak{%
l}_{r}^{8}}\sum_{k=0}^{5}k\mathrm{s}_{k}e_{r}^{k}\cos ^{k-1}\chi _{p}~.
\label{2PNconsistencyEN}
\end{eqnarray}%
Inserting $\mathfrak{a}_{1}^{PN}$, $\mathfrak{a}_{2}^{PN}$, $\mathfrak{a}%
_{1}^{2PN}$ and $\mathfrak{a}_{2}^{2PN}$, we can simplify with $\sin \chi
_{p}$, then after a long, but straightforward calculation we obtain a rank 6
polynomial in $\cos \chi _{p}$, the coefficients of which have to vanish one
by one, as discussed in Appendix \ref{EnJ2PN}.

\subsubsection{The angular momentum conditions}

With the method for verifying the consistency shown in detail above, we can
proceed to verify the consistency of the other constraints.

For the nonspinning 2PN evolution $\alpha =0=\mathfrak{a}\cdot \mathbf{\hat{L%
}}_{\mathbf{N}}$, hence $\left( \mathfrak{J}_{\mathbf{\hat{l}}}\mathbf{,}%
\mathfrak{J}_{\mathbf{\hat{m}}}\mathbf{,}\mathfrak{J}_{\mathbf{\hat{L}}_{%
\mathbf{N}}}\right) =\left( 0,0,\mathfrak{J}\right) $ and the consistency
conditions (\ref{consJdotl})-(\ref{consJdotLN}) simply state that all
components of the dimensionless total angular momentum vector should be
conserved independently (there is no precession involved). The time
derivative of the nontrivial component gives (the same equation emerges by
taking the time derivative of Eq. (\ref{constr3}) with $\kappa _{i}=\alpha
=0 $): 
\begin{equation}
0=\frac{\mathfrak{\dot{l}}_{r}}{\mathfrak{l}_{r}}\left( 1+\epsilon
_{PN}+\epsilon _{2PN}\right) +\dot{\epsilon}_{PN}+\dot{\epsilon}_{2PN}\ .
\label{Jconstr}
\end{equation}%
Following the steps of the proof of consistency of the energy constraint we
obtain, to 1PN\ order accuracy 
\begin{eqnarray}
0 &=&\frac{2\left( 2-\eta \right) e_{r}\left( 1+e_{r}\cos \chi _{p}\right)
^{3}\sin \chi _{p}}{\mathfrak{l}_{r}^{6}}  \notag \\
&&+\mathfrak{a}_{1}^{PN}\sin \chi _{p}-\mathfrak{a}_{2}^{PN}\cos \chi _{p}\ ,
\end{eqnarray}%
then%
\begin{eqnarray}
0 &=&\sum_{k=0}^{3}\mathrm{c}_{1\left( k\right) }^{PN}\cos ^{k}\chi
_{p}-\sum_{k=0}^{2}\mathrm{c}_{2\left( k\right) }^{PN}\cos ^{k+1}\chi _{p} 
\notag \\
&&+2\left( 2-\eta \right) e_{r}\left( 1+e_{r}\cos \chi _{p}\right) \ .
\end{eqnarray}%
This again holds true in each polynomial rank of $\cos \chi _{p}$,
confirming 1PN level consistency of the total angular momentum constraint.

At 2PN order accuracy Eq. (\ref{Jconstr}) gives%
\begin{eqnarray}
0 &=&-\left( \mathfrak{a}_{1}^{2PN}+\frac{\mathfrak{a}_{1}^{PN}}{2\mathfrak{l%
}_{r}^{2}}\mathfrak{b}_{1}\right) \sin \chi _{p}+\mathfrak{a}_{2}^{2PN}\cos
\chi _{p}+\frac{\mathfrak{a}_{2}^{PN}}{2\mathfrak{l}_{r}^{2}}\mathfrak{b}_{2}
\notag \\
&&-\frac{\left( 1+e_{r}\cos \chi _{p}\right) ^{3}\sin \chi _{p}}{8\mathfrak{l%
}_{r}^{8}}\sum_{k=0}^{3}k\mathrm{p}_{k}e_{r}^{k}\cos ^{k-1}\chi _{p}~,
\end{eqnarray}%
with the notations%
\begin{eqnarray}
\mathfrak{b}_{1} &=&\sum_{l=0}^{1}\mathrm{b}_{1\left( l\right) }\cos
^{l}\chi _{p}~,  \notag \\
\mathrm{b}_{1\left( 0\right) } &=&9-7\eta +\left( 1-3\eta \right) e_{r}^{2}~,
\notag \\
\mathrm{b}_{1\left( 1\right) } &=&10\left( 1-\eta \right) e_{r}~,
\end{eqnarray}%
\begin{eqnarray}
\mathfrak{b}_{2} &=&\sum_{l=0}^{2}\mathrm{b}_{2\left( l\right) }\cos
^{l}\chi _{p}~,  \notag \\
\mathrm{b}_{2\left( 0\right) } &=&2\left( 1-\allowbreak 3\eta \right) e_{r}~,
\notag \\
\mathrm{b}_{2\left( 1\right) } &=&\mathrm{b}_{1\left( 0\right) }+e_{r}%
\mathrm{b}_{2\left( 0\right) }~,  \notag \\
\mathrm{b}_{2\left( 2\right) } &=&\mathrm{b}_{1\left( 1\right) }~.
\end{eqnarray}%
By inserting the coefficients, simplifying with $\sin \chi _{p}$ we obtain a
fifth order polynomial in $\cos \chi _{p}$:%
\begin{eqnarray}
0 &=&-2\sum_{k=0}^{5}\mathrm{c}_{1\left( k\right) }^{2PN}\cos ^{k}\chi
_{p}+2\sum_{k=0}^{4}\mathrm{c}_{2\left( k\right) }^{2PN}\cos ^{k+1}\chi _{p}
\notag \\
&&+\left( \sum_{l=0}^{2}\sum_{k=0}^{2}\mathrm{b}_{2\left( l\right) }\mathrm{c%
}_{2\left( k\right) }^{PN}-\sum_{l=0}^{1}\sum_{k=0}^{3}\mathrm{b}_{1\left(
l\right) }\mathrm{c}_{1\left( k\right) }^{PN}\right) \cos ^{k+l}\chi _{p} 
\notag \\
&&-\frac{\left( 1+e_{r}\cos \chi _{p}\right) }{4}\sum_{k=0}^{3}k\mathrm{p}%
_{k}e_{r}^{k}\cos ^{k-1}\chi _{p}~,  \label{2PNconsistencyJ}
\end{eqnarray}%
the coefficients of which can be verified to vanish one by one, as indicated
in Appendix \ref{EnJ2PN}.

Therefore we fulfilled the task to prove the consistency of the nonspinning
evolution and constraint equations up to 2PN accuracy.

\subsection{Consistency of spin and SO contributions}

In the Newton-Wigner-Price SSC the total energy does not contain SO
contributions, therefore the time derivative of Eq. (\ref{constren}) will
not led to any constraints on the leading SO part of the dynamics.

In order to proceed with the consistency of the total angular momentum
constraints, by including the contributions linear in the spin, we need to
remember that $\mathfrak{l}_{r}^{-2}$ represents one relative PN order. The
rhs of the constraints (\ref{constr1})-(\ref{constr2}) contain projections
of the spin and of $\mathbf{L}_{\mathbf{SO}}^{NWP}$, which are linear in the
spins. We will consider only contributions linear in the spins and to
leading order in $\mathfrak{l}_{r}^{-2}$.

The consistency condition of the constraint (\ref{constr1}), given by Eq. (%
\ref{consJdotl}) is 
\begin{align}
0& =\sum_{i=1}^{2}\chi _{i}\frac{d}{d\mathfrak{t}}\Bigl\{\sin \kappa _{i}%
\Bigl[\nu ^{2i-3}\cos \psi _{i}  \notag \\
& -\frac{\eta }{2\mathfrak{l}_{r}^{2}}\left( 4\nu ^{2i-3}+3\right) \left(
1+e_{r}\cos \chi _{p}\right)  \notag \\
& \times \sin \left( \chi _{p}+\psi _{p}\right) \sin \left( \chi _{p}+\psi
_{p}-\psi _{i}\right) \Bigr]\Bigr\}\ .  \label{consSO1}
\end{align}%
From among the time derivatives we explore that $\dot{\chi}_{p}$ has a
Newtonian part $\dot{\chi}_{p}^{N}=\mathcal{O}\left( \mathfrak{l}%
_{r}^{-3}\right) $; then $\mathfrak{\dot{l}}_{r}=\mathcal{O}\left( \mathfrak{%
l}_{r}^{-4}\right) $, $\dot{e}_{r}=\mathcal{O}\left( \mathfrak{l}%
_{r}^{-5}\right) $ and $\dot{\psi}_{p}^{PN}=\mathcal{O}\left( \mathfrak{l}%
_{r}^{-5}\right) $, respectively. We also need to keep in mind that the spin
terms appearing in Eq. (\ref{lrconstr}) combine to $\mathfrak{l}_{r}$. Hence
in Eq. (\ref{consSO1}) we will take into account the leading order $\mathcal{%
O}\left( \mathfrak{l}_{r}^{-5}\right) $ terms, but also those of $\mathcal{O}%
\left( \mathfrak{l}_{r}^{-6}\right) $ which could combine to $\mathcal{O}%
\left( \mathfrak{l}_{r}^{-5}\right) $ terms by virtue of Eq. (\ref{lrconstr}%
). 
\begin{gather}
0=\sum_{i=1}^{2}\dot{\kappa}_{i}^{SO}\nu ^{2i-3}\chi _{i}\cos \kappa
_{i}\cos \psi _{i}-\sum_{i=1}^{2}\chi _{i}\sin \kappa _{i}  \notag \\
\times \Bigl\{\left( \dot{\psi}_{i}^{PN}+\dot{\psi}_{i}^{SO}\right) \nu
^{2i-3}\sin \psi _{i}+\dot{\chi}_{p}^{N}\frac{\eta }{2\mathfrak{l}_{r}^{2}}%
\left( 4\nu ^{2i-3}+3\right)  \notag \\
\times \frac{d}{d\chi _{p}}\left[ \left( 1+e_{r}\cos \chi _{p}\right) \sin
\left( \chi _{p}+\psi _{p}\right) \right.  \notag \\
\times \left. \sin \left( \chi _{p}+\psi _{p}-\psi _{i}\right) \right] %
\Bigr\}\ .
\end{gather}%
To the required order the derivatives are%
\begin{equation}
\dot{\chi}_{p}^{N}=\frac{\left( 1+e_{r}\cos \chi _{p}\right) ^{2}}{\mathfrak{%
l}_{r}^{3}}~,
\end{equation}%
\begin{eqnarray}
~\dot{\kappa}_{i}^{SO} &=&-\frac{{\eta }}{4\mathfrak{l}_{r}^{6}}\left(
1+e_{r}\cos \chi _{p}\right) ^{2}\sin \left( \chi _{p}+\psi _{p}-\psi
_{i}\right)  \notag \\
&&\times \sum_{k=1}^{2}\left( 4\nu ^{2k-3}+3\right) \chi _{k}\sin \kappa
_{k} \left[ e_{r}\allowbreak \cos \left( \psi _{k}-\psi _{p}\right) \right. 
\notag \\
&&+4\cos \left( \chi _{p}+\psi _{p}-\psi _{k}\right)  \notag \\
&&\left. +3e_{r}\cos \left( 2\chi _{p}+\psi _{p}-\psi _{k}\right) \right] ~,
\end{eqnarray}%
\begin{eqnarray}
\dot{\psi}_{i}^{SO} &=&\frac{\eta }{2\mathfrak{l}_{r}^{5}}\left( 1+e_{r}\cos
\chi _{p}\right) ^{3}\left( 4+3\nu ^{3-2i}\right)  \notag \\
&&-\frac{\eta }{4\mathfrak{l}_{r}^{6}}\left( 1+e_{r}\cos \chi _{p}\right)
^{2}  \notag \\
&&\times \left[ \sin \left( \chi _{p}+\psi _{p}\right) \cot \alpha -\cos
\left( \chi _{p}+\psi _{p}-\psi _{i}\right) \cot \kappa _{i}\right]  \notag
\\
&&\times \sum_{k=1}^{2}\left( 4\nu ^{2k-3}+3\right) \chi _{k}\sin \kappa
_{k} \left[ e_{r}\allowbreak \cos \left( \psi _{k}-\psi _{p}\right) \right. 
\notag \\
&&+4\cos \left( \chi _{p}+\psi _{p}-\psi _{k}\right)  \notag \\
&&\left. +3e_{r}\cos \left( 2\chi _{p}+\psi _{p}-\psi _{k}\right) \right] ~.
\end{eqnarray}%
We get%
\begin{gather}
0=\sum_{i=1}^{2}\chi _{i}\sin \kappa _{i}\left( 4\nu ^{2i-3}+3\right) \Bigl\{%
\left( 1+e_{r}\cos \chi _{p}\right) \sin \psi _{i}  \notag \\
+\frac{d}{d\chi _{p}}\left[ \left( 1+e_{r}\cos \chi _{p}\right) \sin \left(
\chi _{p}+\psi _{p}\right) \sin \left( \chi _{p}+\psi _{p}-\psi _{i}\right) %
\right] \Bigr\}  \notag \\
-\frac{1}{2\mathfrak{l}_{r}}\sum_{i=1}^{2}\nu ^{2i-3}\chi _{i}\left[ \sin
\kappa _{i}\sin \psi _{i}\cot \alpha -\cos \kappa _{i}\sin \left( \chi
_{p}+\psi _{p}\right) \right]  \notag \\
\times \sin \left( \chi _{p}+\psi _{p}\right) \sum_{k=1}^{2}\left( 4\nu
^{2k-3}+3\right) \chi _{k}\sin \kappa _{k}  \notag \\
\times \left[ e_{r}\allowbreak \cos \left( \psi _{k}-\psi _{p}\right) +4\cos
\left( \chi _{p}+\psi _{p}-\psi _{k}\right) \right.  \notag \\
\left. +3e_{r}\cos \left( 2\chi _{p}+\psi _{p}-\psi _{k}\right) \right] ~,
\end{gather}%
which, after exploring the constraint (\ref{lrconstr}) in the third line,
becomes%
\begin{gather}
0=\sum_{i=1}^{2}\chi _{i}\sin \kappa _{i}\left( 4\nu ^{2i-3}+3\right) \Bigl\{%
\left( 1+e_{r}\cos \chi _{p}\right) \sin \psi _{i}  \notag \\
+\frac{d}{d\chi _{p}}\left[ \left( 1+e_{r}\cos \chi _{p}\right) \sin \left(
\chi _{p}+\psi _{p}\right) \sin \left( \chi _{p}+\psi _{p}-\psi _{i}\right) %
\right] \Bigr\}  \notag \\
-\frac{1}{2}\sin \left( \chi _{p}+\psi _{p}\right) \sum_{k=1}^{2}\left( 4\nu
^{2k-3}+3\right) \chi _{k}\sin \kappa _{k}  \notag \\
\times \left[ e_{r}\allowbreak \cos \left( \psi _{k}-\psi _{p}\right) +4\cos
\left( \chi _{p}+\psi _{p}-\psi _{k}\right) \right.  \notag \\
\left. +3e_{r}\cos \left( 2\chi _{p}+\psi _{p}-\psi _{k}\right) \right] ~.
\end{gather}%
This can be shown to identically hold, hence to leading order in the SO
contributions the consistency condition (\ref{consJdotl}) is obeyed.

For the consistency condition (\ref{consJdotm}), we calculate the derivative 
$\mathfrak{\dot{J}}_{\mathbf{\hat{m}}}$ as the rhs of Eq. (\ref{constr2}).
We count the orders at which the dimensionless variables change, obtaining
to leading order%
\begin{gather}
\mathfrak{\dot{J}}_{\mathbf{\hat{m}}}=\sum_{i=1}^{2}\chi _{i}\sin \kappa _{i}%
\Bigl[\dot{\psi}_{i}^{PN}\nu ^{2i-3}\cos \psi _{i}+\dot{\chi}_{p}^{N}\frac{%
\eta \left( 4\nu ^{2i-3}+3\right) }{2\mathfrak{l}_{r}^{2}}  \notag \\
\times \frac{d}{d\chi _{p}}\left[ \left( 1+e_{r}\cos \chi _{p}\right) \cos
\left( \chi _{p}+\psi _{p}\right) \sin \left( \chi _{p}-\zeta _{i}\right) %
\right] \Bigr]\ ,
\end{gather}%
or by inserting the respective time evolutions 
\begin{gather}
\mathfrak{\dot{J}}_{\mathbf{\hat{m}}}=\frac{\eta \left( 1+e_{r}\cos \chi
_{p}\right) ^{2}}{2\mathfrak{l}_{r}^{5}}\sum_{i=1}^{2}\left( 4\nu
^{2i-3}+3\right) \chi _{i}\sin \kappa _{i}  \notag \\
\times \Bigl[\frac{d}{d\chi _{p}}\left[ \left( 1+e_{r}\cos \chi _{p}\right)
\cos \left( \chi _{p}+\psi _{p}\right) \sin \left( \chi _{p}-\zeta
_{i}\right) \right]  \notag \\
+\left( 1+e_{r}\cos \chi _{p}\right) \cos \left( \psi _{p}+\zeta _{i}\right) %
\Bigr]\ ,
\end{gather}%
Then as $\mathfrak{a}_{3}^{SO}=\mathcal{O}\left( \mathfrak{l}%
_{r}^{-7}\right) $ and to leading order $\mathfrak{J}\cos \alpha =\mathfrak{l%
}_{r}$, the second term in Eq. (\ref{consJdotm}) is also $\mathcal{O}\left( 
\mathfrak{l}_{r}^{-5}\right) $. As the spin magnitudes are arbitrary
constants, to leading order the consistency condition (\ref{consJdotm})
splits into two equations, one for each spin direction:%
\begin{gather}
0=2\frac{d}{d\chi _{p}}\left[ \left( 1+e_{r}\cos \chi _{p}\right) \cos
\left( \chi _{p}+\psi _{p}\right) \sin \left( \chi _{p}-\zeta _{i}\right) %
\right]  \notag \\
+2\left( 1+e_{r}\cos \chi _{p}\right) \cos \left( \psi _{p}+\zeta
_{i}\right) -\cos \left( \psi _{p}+\chi _{p}\right)  \notag \\
\times \left[ e_{r}\allowbreak \cos \zeta _{i}+4\cos \left( \chi _{p}-\zeta
_{i}\right) +3e_{r}\cos \left( 2\chi _{p}-\zeta _{i}\right) \right] ~.
\end{gather}%
This can be shown to identically hold, hence to leading order in the SO
contributions the consistency condition (\ref{consJdotm}) is obeyed.

Finally for the consistency condition (\ref{consJdotLN}) we calculate the
derivative $\mathfrak{\dot{J}}_{\mathbf{\hat{L}}_{\mathbf{N}}}$ as the rhs
of Eq. (\ref{constr3}), with only the Newtonian and SO terms included. We
again explore the orders at which the dimensionless variables change,
obtaining 
\begin{equation}
\mathfrak{\dot{J}}_{\mathbf{\hat{L}}_{\mathbf{N}}}=\mathfrak{\dot{l}}%
_{r}^{SO}+\dot{\chi}_{p}^{N}\frac{\eta e_{r}\sin \chi _{p}}{2\mathfrak{l}%
_{r}^{2}}\sum_{i=1}^{2}\left( 4\nu ^{2i-3}+3\right) \chi _{i}\cos \kappa
_{i}=0\ .  \label{consJdotLN1}
\end{equation}%
The second identity emerges by inserting the explicit expressions of $%
\mathfrak{\dot{l}}_{r}^{SO}$, $\mathfrak{a}_{1}^{SO}$, $\mathfrak{a}%
_{2}^{SO} $ and $\dot{\chi}_{p}^{N}$ and holds at $\mathcal{O}\left( 
\mathfrak{l}_{r}^{-5}\right) $. Then as $\mathfrak{a}_{3}^{SO}=\mathcal{O}%
\left( \mathfrak{l}_{r}^{-7}\right) $ and $\mathfrak{J}\sin \alpha =\mathcal{%
O}\left( 1\right) $, the second term in Eq. (\ref{consJdotLN}) is $\mathcal{O%
}\left( \mathfrak{l}_{r}^{-6}\right) $, to be dropped. Hence to leading
order in the SO contributions this last consistency condition is also obeyed.

Remarkably by Eq. (\ref{consJdotLN1}) we have proven that in the basis $%
\left( \mathbf{\hat{l},\hat{m},\hat{L}}_{\mathbf{N}}\right) $ not only $%
\mathfrak{J}_{\mathbf{\hat{l}}}$, but also $\mathfrak{J}_{\mathbf{\hat{L}}_{%
\mathbf{N}}}$ is conserved to leading order. This indicates how well this
precessing basis is adapted to the dynamics of the binary.

\subsection{Consistency of SS contributions}

The time derivative of the Keplerian + SS part of the energy constraint (\ref%
{constren}) gives%
\begin{eqnarray}
0 &=&\mathfrak{a}_{1}^{SS}\sin \chi _{p}-\mathfrak{a}_{2}^{SS}\allowbreak
\left( e_{r}+\cos \chi _{p}\right)  \notag \\
&&-\frac{3\eta \left( 1+e_{r}\cos \chi _{p}\right) ^{4}}{2\mathfrak{l}%
_{r}^{8}}\chi _{1}\chi _{2}  \notag \\
&&\times \biggl\{2e_{r}\cos \kappa _{1}\cos \kappa _{2}\sin \chi _{p}-\sin
\kappa _{1}\sin \kappa _{2}  \notag \\
&&\times \left\{ e_{r}\sin \chi _{p}\left[ \cos \zeta _{\left( -\right)
}+3\cos \left( 2\chi _{p}-\zeta _{\left( +\right) }\right) \right] \right. 
\notag \\
&&\left. +2\left( 1+e_{r}\cos \chi _{p}\right) \sin \left( 2\chi _{p}-\zeta
_{\left( +\right) }\right) \right\} \biggr\}~.
\end{eqnarray}%
After inserting the dimensionless perturbing force components, it is not
difficult to verify that the coefficients of $\cos \kappa _{1}\cos \kappa
_{2}$ and $e_{r}^{k}\sin \kappa _{1}\sin \kappa _{2}\,$\ (with $k=0,1,2$)
all vanish, therefore the above equation is an identity.

The total angular momentum does not contain SS contributions, therefore the
time derivatives of Eqs. (\ref{constr1})-(\ref{constr3}) do not impose any
constraints at the leading SS order of the dynamics.

\subsection{Consistency of QM contributions}

The time derivative of the Keplerian + QM part of the energy constraint (\ref%
{constren}) gives the QM\ order equation:%
\begin{eqnarray}
0 &=&\mathfrak{a}_{1}^{QM}\sin \chi _{p}-\mathfrak{a}_{2}^{QM}\allowbreak
\left( e_{r}+\cos \chi _{p}\right)  \notag \\
&&+\frac{3\eta \left( 1+e_{r}\cos \chi _{p}\right) ^{4}}{2\mathfrak{l}%
_{r}^{8}}\sum_{i=1}^{2}w_{i}\chi _{i}^{2}\nu ^{2i-3}  \notag \\
&&\times \left\{ 2\left( 1+e_{r}\cos \chi _{p}\right) \sin ^{2}\!\kappa
_{i}\cos \left( \!\chi \!_{p}-\zeta _{i}\right) \sin \left( \!\chi
\!_{p}-\zeta _{i}\right) \right.  \notag \\
&&\left. -e_{r}\sin \chi _{p}\left[ 1-\!3\!\sin ^{2}\!\kappa _{i}\cos
^{2}\!\left( \!\chi \!_{p}-\zeta _{i}\right) \right] \right\} ~.
\end{eqnarray}%
After inserting the dimensionless perturbing force components, it is not
difficult to verify that the coefficients of $\sin ^{0}\kappa _{i}$ and $%
e_{r}^{k}\sin ^{2}\kappa _{i}\,$\ (with $i=1,2$ and $k=0,1,2$) all vanish,
therefore the above equation is an identity.

The total angular momentum does not contain QM contributions, therefore the
time derivatives of Eqs. (\ref{constr1})-(\ref{constr3}) do not impose any
constraints at the leading QM order of the dynamics.

\section{Chameleon orbits\label{eccentrichyperbolic}}

In this section we investigate highly eccentric orbits, with $e_{r}\approx 1$%
. Such orbits could be induced by three-body interactions, also could arise
in the central regions of galaxies. Stellar orbits in these regions were
already investigated in order to test the spin of the central supermassive
black hole \cite{MerrittAlexanderMikkolaWill}. Gravitational radiation from
such highly eccentric orbits was recently discussed in Refs. \cite%
{Kennefick,zoomwhirlGeod,zoomwhirlNum,Capture,zoomwhirlPN,Vasuth}.

Our aim here is to apply the equations we derived for the study of
conservative dynamics in order to test general relativistic features of
gravity. Indeed, it is well known (from example from the general
relativistic Oppenheimer-Volkoff equation) that general relativity predicts
stronger gravity at short distances, than arising in the Newtonian
description. Therefore we expect that for sufficiently large values of the
PN parameter the highly eccentric orbits could produce the following
feature. Orbits which (in terms of the eccentricity $e_{r}$ of the
osculating orbit) are hyperbolic, could become elliptic close to the
periastron. Another way to see that is due the potential well deepening
faster in general relativity than in Newtonian gravity. Such orbits locally
look hyperbolic at large distances and elliptic at short distances. Hence we
call them chameleon orbits.

It has been known earlier \cite{Capture} that due to gravitational radiation
hyperbolic orbits can turn into elliptic. Our analysis shows a similar
effect already at the conservative level. We were able to illustrate this
behavior already by including the 1PN corrections to the Keplerian dynamics,
by evolving numerically the system of equations (\ref{lrdot})-(\ref{erdot})
and (\ref{chipdot}). For this case of zero spins ($\chi _{1}=\chi _{2}=0$)
the system of differential equations is closed. The chameleon behavior is
presented on Fig. \ref{Fig1}, both for equal mass binaries $\nu =1$ (left
panel) and for a highly asymmetric system, with mass ratio $\nu =1/30$
(right panel). The initial values were chosen at the periastron as $%
e_{r}\left( \chi _{p}=0\right) =0.96$\ and $\varepsilon \left( \chi
_{p}=0\right) =Gm/c^{2}r_{\min }=0.01$\ in both cases. Then $l_{r}\left(
\chi _{p}=0\right) $\ is derived from (\ref{lrrmin}). The function $%
e_{r}\left( \chi _{p}\right) $ is symmetric to the periastron and its
asymptotic values are larger for decreasing mass ratios. The orbits $R\cos
\chi _{p}$ vs. $R\sin \chi _{p}$\ with $R=c^{2}r/Gm$ are represented by the
green curve on Fig. \ref{Fig1}. The domains with $e_{r}<1$ and $e_{r}>1$ are
also indicated.

We then proceeded to study the modifications induced by the spins on these
orbits. For this we supplemented the 1PN corrections with the leading order
SO contribution. For simplicity we have chosen non-precessing
configurations, with the spins of the components either aligned or
anti-aligned with the orbital momentum. In this case the angles $\kappa _{1}$
and $\kappa _{2}$\ remain constants during the motion and the system of
equations (\ref{lrdot})-(\ref{erdot}) and (\ref{chipdot}) is again closed.
The orientations of the orbital angular momentum and spin vectors are
indicated by arrows on the panels. Both dimensionless spin parameters are
taken as $\chi _{i}=0.9982$, which is the canonical spin limit, achieved by
black holes with radiating accretion disks leading to photon capture \cite%
{Thorne74}. The initial conditions were the same as for the chameleon orbits
represented on Fig. \ref{Fig1}. For anti-parallel spins the two SO
contributions cancel in the equations, therefore the orbit is identical to
the one represented on the left panel of Fig. \ref{Fig1}. When the spins are
parallel, the orbits become asymmetric with respect to the periastron, as
shown on Fig. \ref{Fig2}. Then a further distinction comes from the
alignment or anti/alignment of the spins with the orbital angular momentum.
In the anti-aligned case (left panel), the asymptotic value of $e_{r}\left(
\chi _{p}\right) $ is larger before the periastron than after it. For spins
aligned with $\mathbf{L}_{\mathbf{N}}$\ (right panel) the evolution of $%
e_{r} $\ shows an opposite trend, also the difference between the asymptotic
values becomes slightly smaller.

On Fig. \ref{Fig3} we show various chameleon orbits due the 1PN and SO
contributions in the equations of motion for unequal mass ($\nu =1/30$)
spinning binaries, again for $\chi _{i}=0.9982$. Each of the spins could be
either aligned or anti-aligned with the orbital angular momentum. The
various possibilities are represented by arrows on the panels of the figure.
When the spins are parallel with each other (upper left and lower right
panels), similar asymmetric evolutions occur than in the case of equal
masses but the difference between the asymptotic values of $e_{r}$ is
enhanced by the small mass ratio. The asymmetric character (e.g. which
asymptotic value of $e_{r}$\ is bigger) of the orbits is determined by the
orientation of the spin of the larger mass with respect to the orbital
angular momentum, as shown on the upper right and lower left panels. The
orientation of the second spin has but little influence on the precise
asymptotic values of $e_{r}$, while the generic shape of the chameleon
evolutions is unaffected, as can be seen by comparing the upper panels or
the lower panels.

\begin{widetext}

\begin{figure}[th]
\includegraphics[height=7.2cm,angle=270]{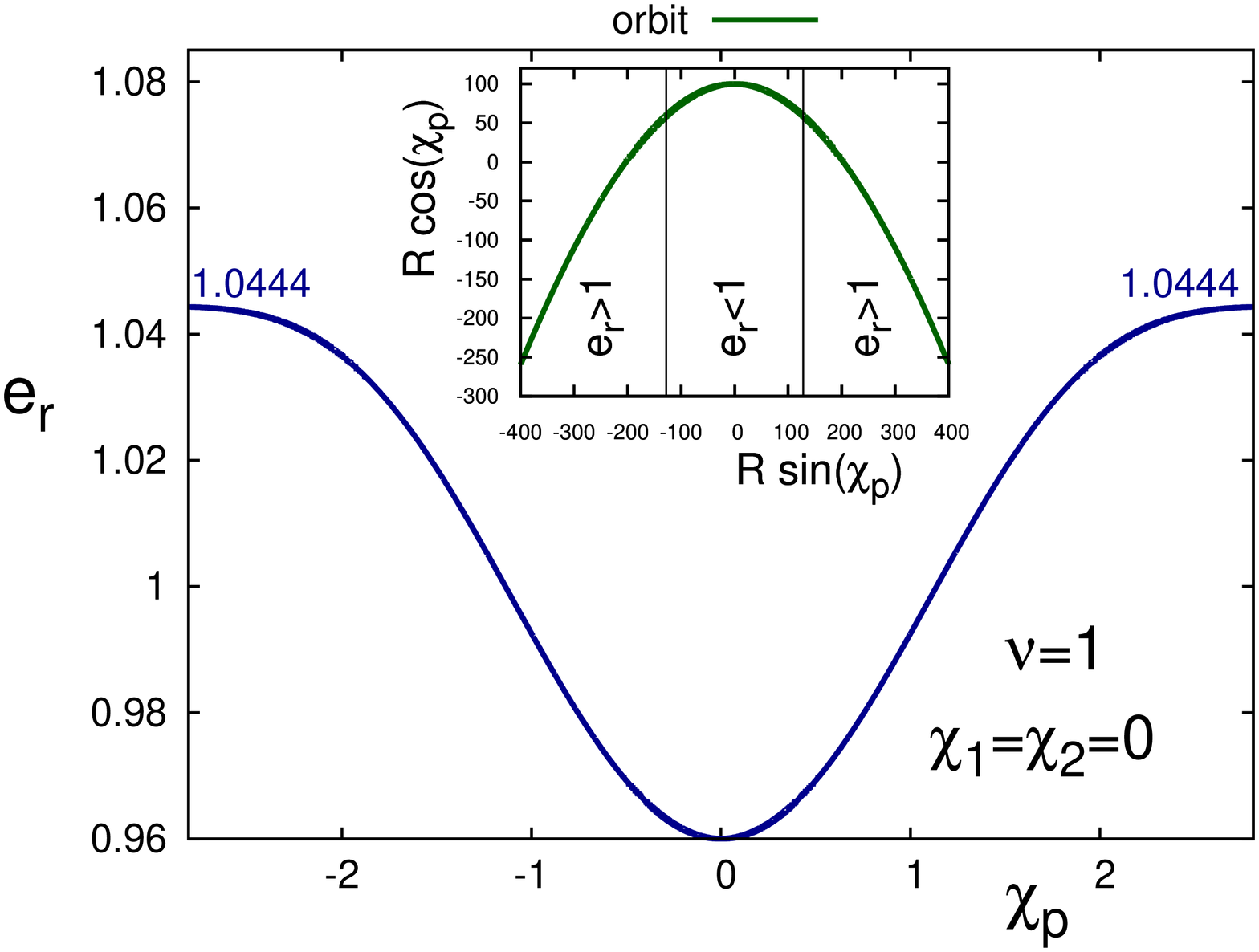} \hskip0.4cm%
\includegraphics[height=7.2cm,angle=270]{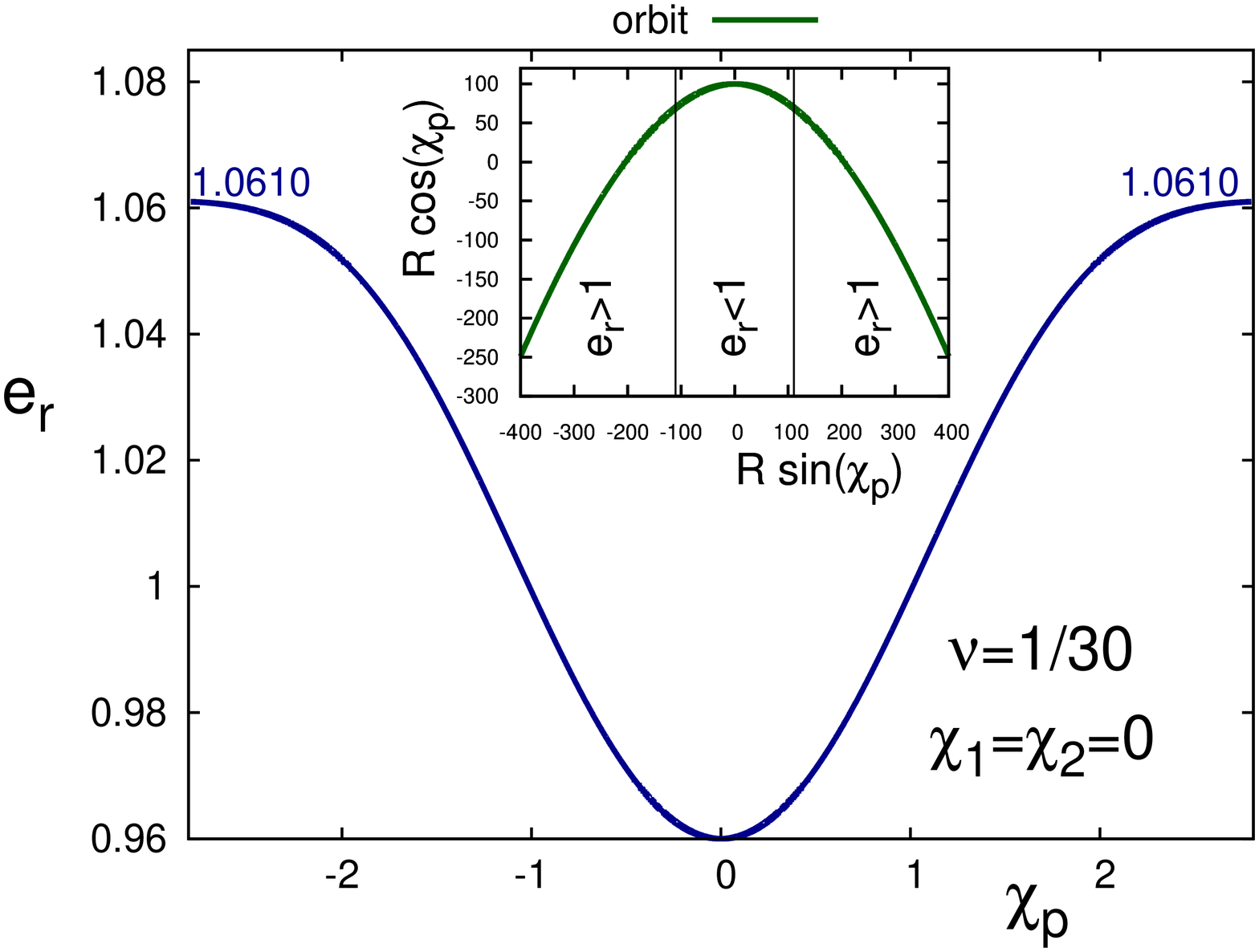} \vskip0.4cm
\caption{Chameleon orbits due to the 1PN order effects are shown for equal ($%
\protect\nu =1$, left panel) and asymmetric ($\protect\nu =1/30$, right
panel) mass binaries. The chameleon behaviour is characterized by the
trespassing of the function $e_{r}\left( \protect\chi _{p}\right) $ across
the value 1 (indicated in blue). Initial conditions are fixed at the
periastron as $e_{r}\left( \protect\chi _{p}=0\right) =0.96$ and\textbf{\ }$%
\protect\varepsilon \left( \protect\chi _{p}=0\right) =Gm/c^{2}r_{\min }=0.01
$. The asymptotic values of $e_{r}$ are given (in blue) on the left and
right sides on each panel. The orbits $R\cos \protect\chi _{p}$ vs. $R\sin 
\protect\chi _{p}$ with\textbf{\ }$R=c^{2}r/Gm$ are shown by the (green)
curve in the smaller boxes. The domains with $e_{r}<1$ and $e_{r}>1$,
respectively are also indicated.}
\label{Fig1}
\end{figure}

\begin{figure}[th]
\includegraphics[height=7.2cm,angle=270]{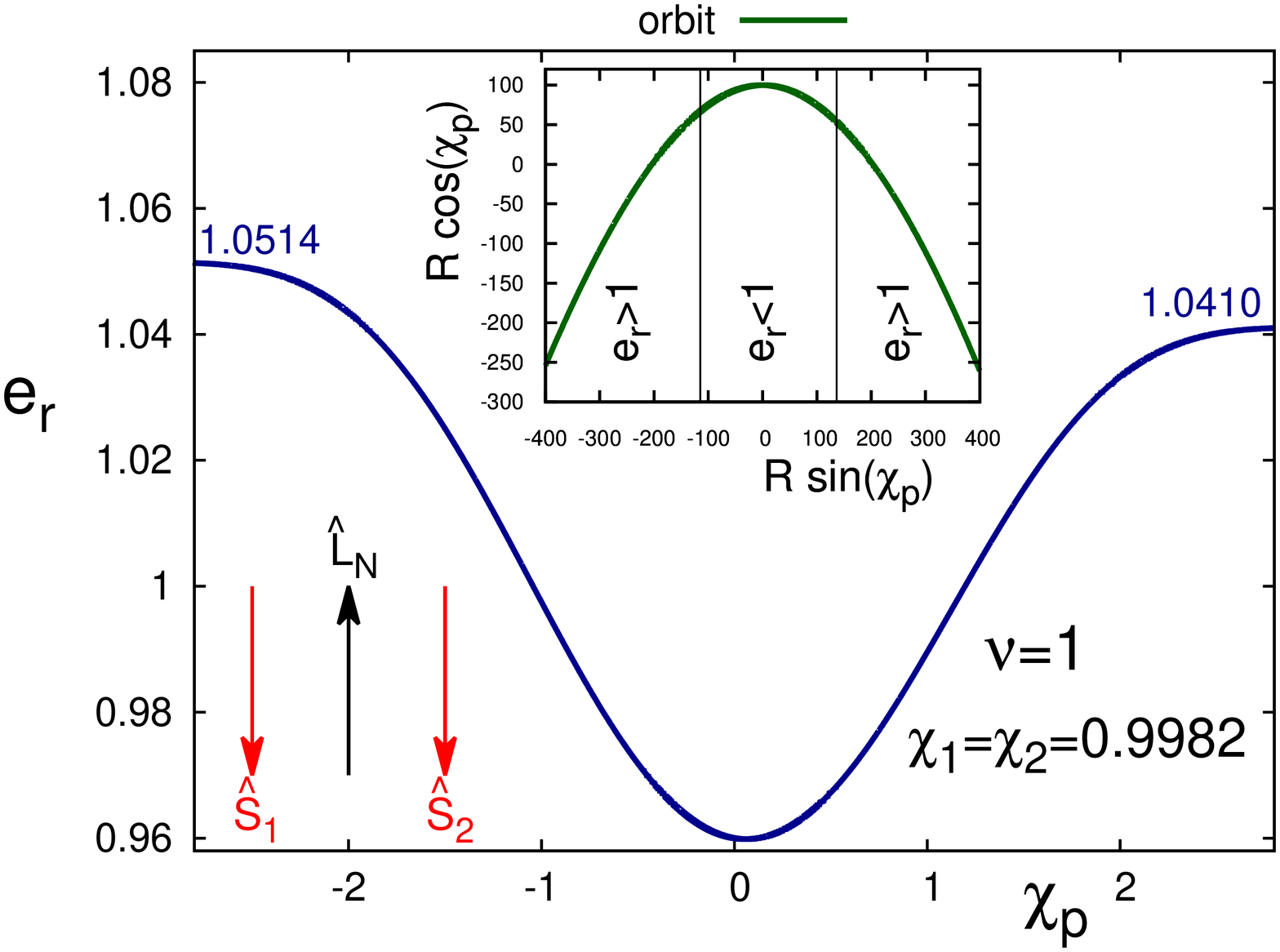} \hskip0.4cm%
\includegraphics[height=7.2cm,angle=270]{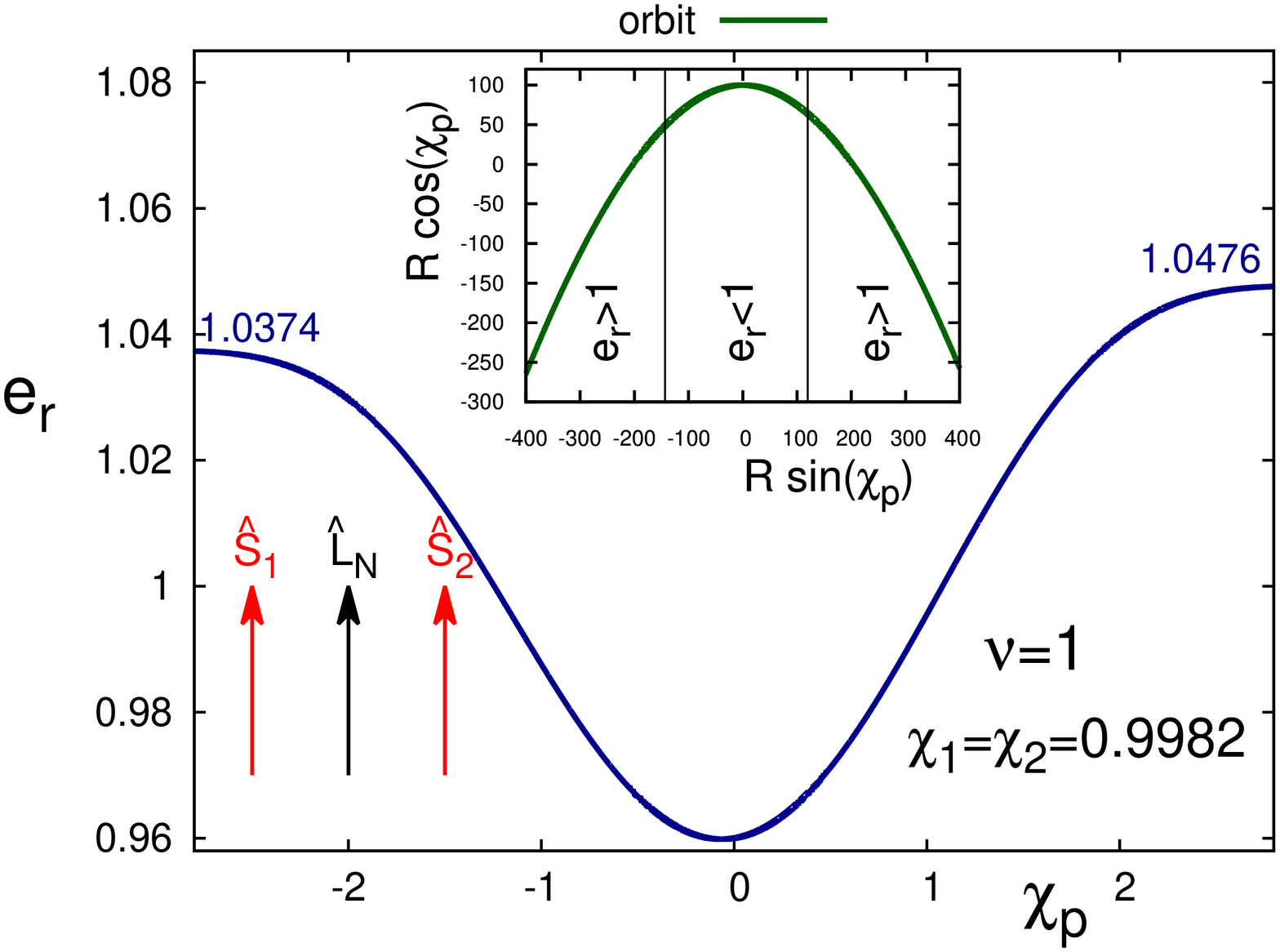} \vskip0.4cm
\caption{Chameleon orbits due to 1PN and SO effects for binaries with equal
masses and spins ($\protect\chi _{1}=\protect\chi _{2}=0.9982$). The curves
and initial conditions are as on Fig. \protect\ref{Fig1}. On the left
(right) panel the spins are anti-aligned (aligned) with the orbital angular
momentum.}
\label{Fig2}
\end{figure}

\begin{figure}[th]
\includegraphics[height=7cm,angle=270]{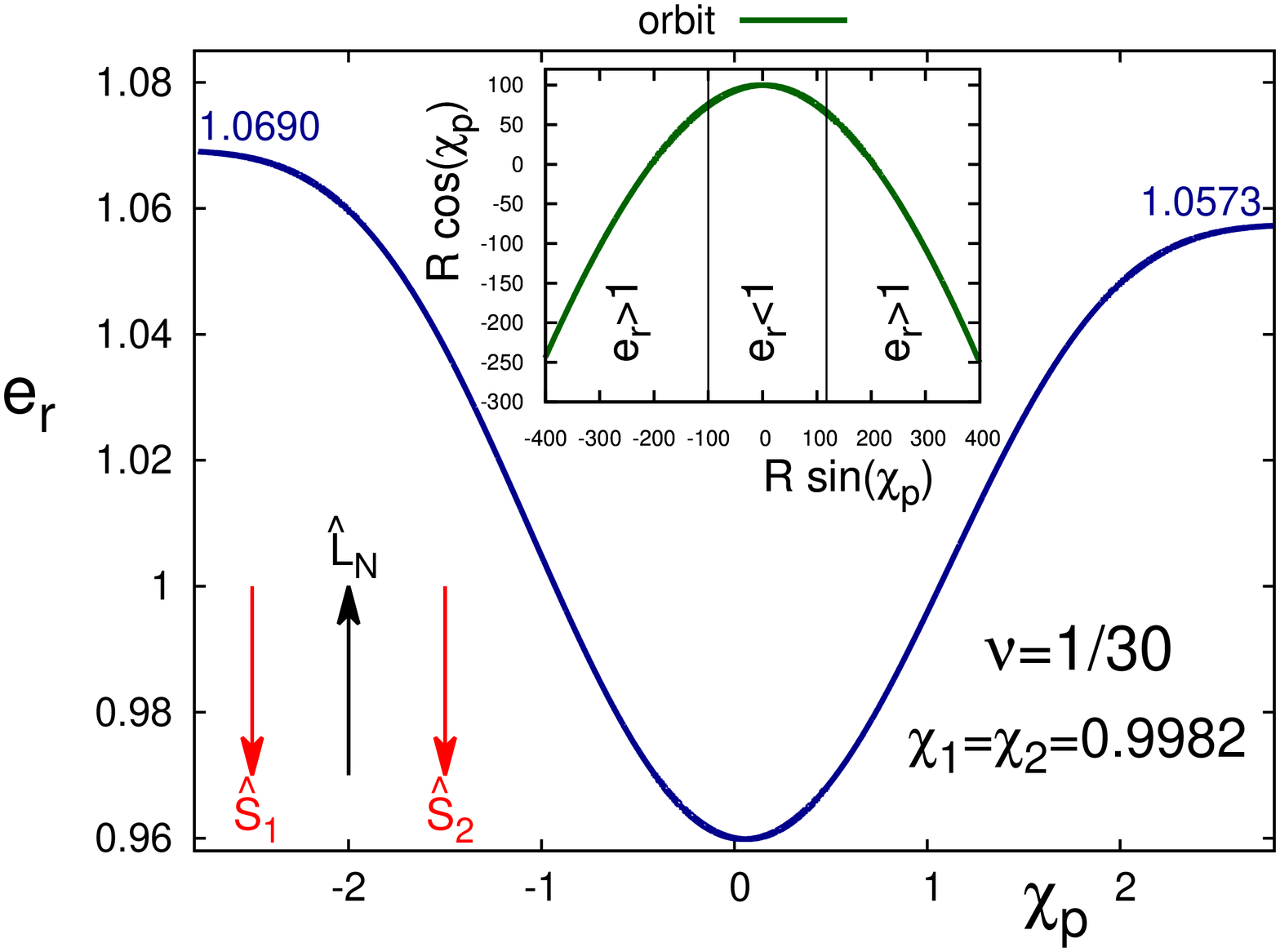} \hskip0.4cm%
\includegraphics[height=7cm,angle=270]{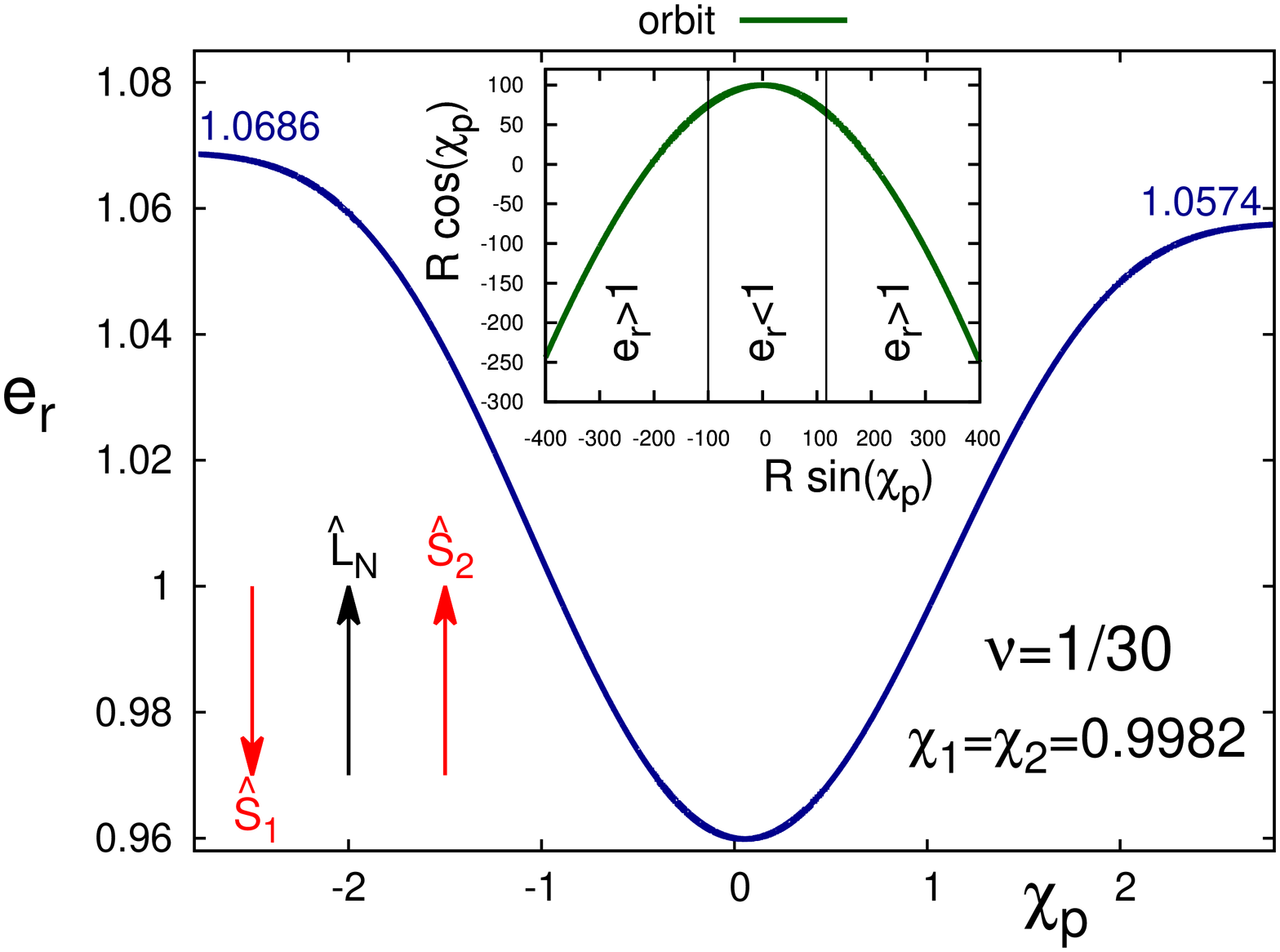} \vskip0.4cm %
\includegraphics[height=7cm,angle=270]{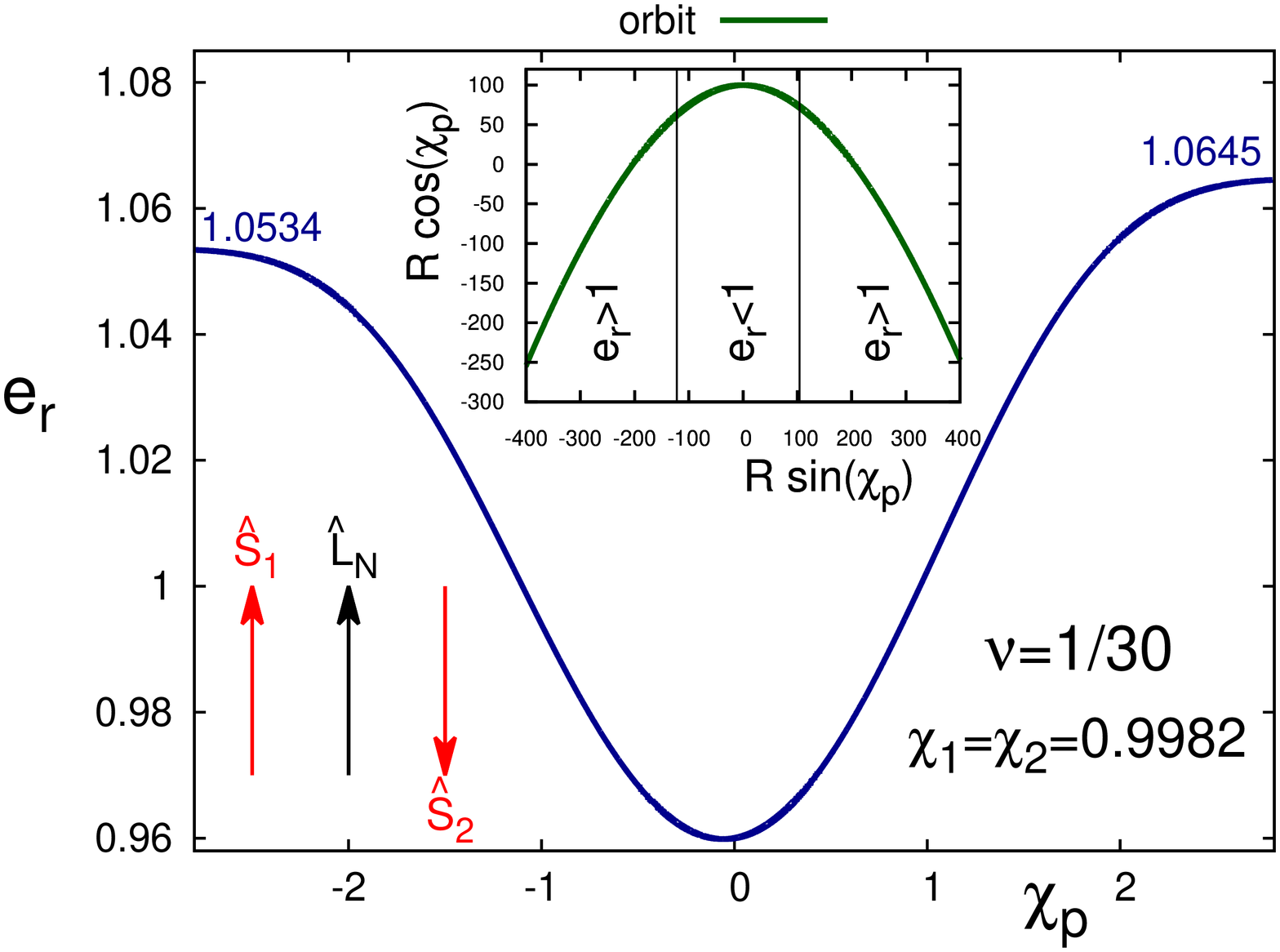} \hskip0.4cm%
\includegraphics[height=7cm,angle=270]{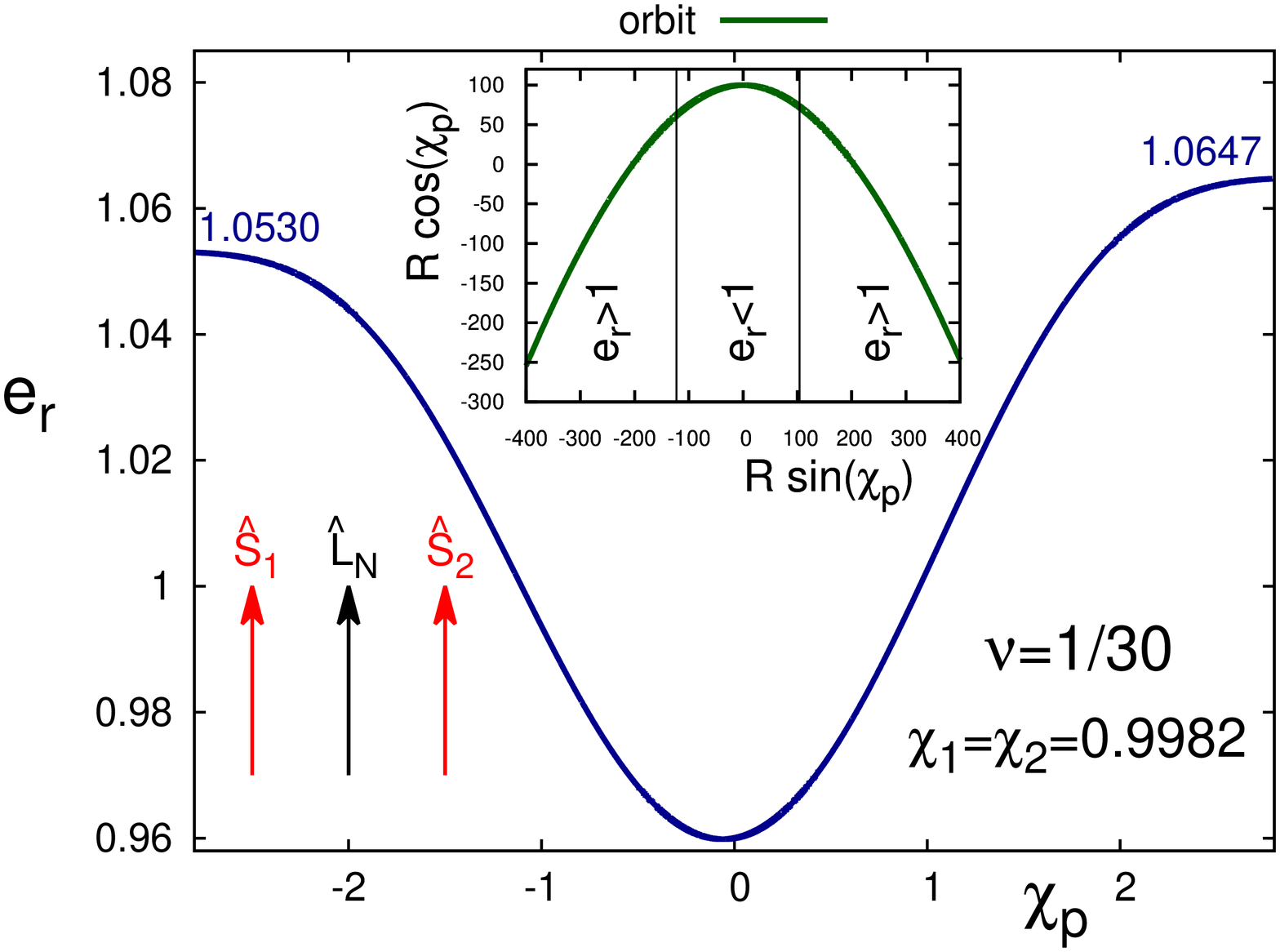} \vskip0.4cm
\caption{Chameleon orbits created by 1PN and SO effects are represented for
unequal mass ($\protect\nu =1/30$) spinning binaries. The same functions as
on Fig. \protect\ref{Fig1} are shown for the same initial conditions. The
relative direction of the spins and the orbital momentum are indicated by
arrows. The dimensionless spin values are the same $\protect\chi _{1}=%
\protect\chi _{2}=0.9982$ in all panels.}
\label{Fig3}
\end{figure}

\end{widetext}

\section{Concluding Remarks\label{CoRe}}

In this paper we have considered the conservative evolution of spinning
compact binaries up to the second post-Newtonian order accuracy, by
including the leading order spin-orbit, spin-spin and mass
quadrupole-monopole contributions. The novel feature of the discussion is
that it had been presented in terms of suitably chosen dimensionless
variables. These are i) the variables replacing the traditional orbital
elements of celestial mechanics: a dimensionless version of the orbital
angular momentum, the eccentricity and three Euler angles characterising the
orientation of the orbit and the orbital plane with respect to the total
angular momentum vector, also ii) dimensionless spin magnitudes (smaller
than one for both black holes and neutron stars) together with the spin
azimuthal and polar angles. The preferred reference system of this analysis
is tied to the orbital angular momentum and periastron.

As a main result we derived a system of first order differential equations
in a compact form, for a set of 9 dimensionless variables encompassing both
the orbital elements and the spin angles (the spin magnitudes being
conserved). These are supplemented by the evolution equation of the true
anomaly, which closes the differential system.

These evolutions are constrained by the conservation laws of energy and
total angular momentum vector holding at 2PN\ order. As required by the
generic theory of constrained dynamical systems we analyzed the consistency
of the constraints, e.g. their compatibility with the evolution equations,
and proved that they are preserved by the evolution.

We applied the formalism to show the existence of orbits with unusual
features. Close to the periastron, the osculating orbits of these
trajectories with eccentricity close to one change from hyperbolic to
elliptic, then back to hyperbolic. Hence these orbits (as characterised by
the eccentricity of their osculating orbit) look open, then closed, then
open again during the passage through the periastron. These \textit{chameleon%
} \textit{orbits} evolve from elliptic (locally, in a Newtonian sense) close
to the periastron into hyperbolic (in the same sense) at large distances.
Such a property emerges due to the fact that General Relativity predicts
stronger gravity (deeper potential wells) at short distances than Newtonian
theory does, as also illustrated by the hydrostatic equilibrium in
relativistic stars.

We analysed the chameleon orbits as function of mass ratios and spin
orientations, for aligned and anti-aligned spin and orbital angular momentum
configurations. Without spin, these orbits are symmetric with respect to the
periastron. The farther the mass ratio is from unity, the larger is the
change in the eccentricity of the osculating orbit, hence the easier to find
such chameleon orbits.

The presence of spins can not be detected when the masses are equal and the
spins anti-aligned with each other. In all other cases with spin, they
induce an asymmetry with respect to the periastron. One aspect of this
asymmetry is that the minimum of the eccentricity is not in the periastron,
as can be seen on Figs. \ref{Fig2}-\ref{Fig3}. As a rule we found that the
alignment of the total spin $\mathbf{S}_{\mathbf{1}}+\mathbf{S}_{2}$ with
the orbital angular momentum shifts the minimum eccentricity point of the
trajectory before the periastron, while the anti-alignment shifts it after
the periastron. These results hold both in the equal mass and in the
asymmetric mass cases.

This feature of relativistic orbits is complementary to how the rotation or
counterrotation of a particle in circular orbit about a rotating black hole
affects the location of the innermost stable orbit. In our case co-rotation
apparently speeds up the (reduced mass) particle, while counter-rotation
slows it down, after leaving the periastron.

\section{Acknowledgements}

L\'{A}G was supported by the European Union and the State of Hungary,
cofinanced by the European Social Fund in the framework of T\'{A}MOP 4.2.4.
A/2-11-/1-2012-0001 `National Excellence Program'. ZK has been supported by
OTKA Grant No. 100216.

\appendix

\section{Computational details for verifying the 2PN accurate, nonspinning
consistency conditions\label{EnJ2PN}}

In this Appendix we give computational details for the proof of the 2PN
accurate consistency conditions in the absence of spins.

First we discuss the energy consistency condition (\ref{2PNconsistencyEN}).
After inserting the acceleration components $\mathfrak{a}_{1}^{PN}$, $%
\mathfrak{a}_{2}^{PN}$, $\mathfrak{a}_{1}^{2PN}$ and $\mathfrak{a}_{2}^{2PN}$%
, then simplifying with $\sin \chi _{p}$ the coefficients of the 6th order
polynomial in $\cos \chi _{p}$, enlisted below, have to vanish.

The terms without $\cos \chi _{p}$ give:%
\begin{eqnarray}
0 &=&\mathrm{s}_{1}e_{r}+16\mathrm{c}_{1\left( 0\right) }^{2PN}-16e_{r}%
\mathrm{c}_{2\left( 0\right) }^{2PN}-2\mathrm{c}_{2\left( 0\right) }^{PN}%
\frac{d\mathrm{q}_{0}}{de_{r}}  \notag \\
&&+2\mathrm{c}_{1\left( 0\right) }^{PN}\left( 2\mathrm{q}_{1}+\frac{d\mathrm{%
q}_{0}}{de_{r}}e_{r}-4\mathrm{q}_{0}\right) ~,
\end{eqnarray}%
which add up to zero after inserting the definitions of the coefficients $%
\mathrm{c}_{i(k)}^{PN}$, $\mathrm{c}_{i(k)}^{2PN}$, $\mathrm{q}_{i}$, $%
\mathrm{s}_{i}$.

The coefficient of $\cos \chi _{p}$ gives:%
\begin{eqnarray}
0 &=&2\mathrm{s}_{2}e_{r}^{2}+\mathrm{s}_{1}e_{r}^{2}+16\left( \mathrm{c}%
_{1\left( 1\right) }^{2PN}-\mathrm{c}_{2\left( 0\right) }^{2PN}\right) 
\notag \\
&&+16e_{r}\mathrm{c}_{1\left( 0\right) }^{2PN}-16e_{r}\mathrm{c}_{2\left(
1\right) }^{2PN}-16\allowbreak e_{r}^{2}\mathrm{c}_{2\left( 0\right) }^{2PN}
\notag \\
&&-2\mathrm{c}_{2\left( 0\right) }^{PN}\left( \frac{d\mathrm{q}_{1}}{de_{r}}+%
\frac{d\mathrm{q}_{0}}{de_{r}}\right) e_{r}-2\mathrm{c}_{2\left( 1\right)
}^{PN}\frac{d\mathrm{q}_{0}}{de_{r}}  \notag \\
&&+2\left( \mathrm{c}_{1\left( 1\right) }^{PN}-\mathrm{c}_{2\left( 0\right)
}^{PN}\right) \left[ 2\left( \mathrm{q}_{1}-2\mathrm{q}_{0}\right) +\frac{d%
\mathrm{q}_{0}}{de_{r}}e_{r}\right]  \notag \\
&&+2\mathrm{c}_{1\left( 0\right) }^{PN}\left[ 2\left( 2\mathrm{q}_{2}-%
\mathrm{q}_{1}\right) e_{r}+\frac{d\mathrm{q}_{1}}{de_{r}}e_{r}^{2}+\frac{d%
\mathrm{q}_{0}}{de_{r}}\right] ~,
\end{eqnarray}%
adding up to zero after inserting the definitions of the coefficients $%
\mathrm{c}_{i(k)}^{PN}$, $\mathrm{c}_{i(k)}^{2PN}$, $\mathrm{q}_{i}$, $%
\mathrm{s}_{i}$.

The coefficient of $\cos ^{2}\chi _{p}$ gives:%
\begin{eqnarray}
0 &=&3\mathrm{s}_{3}e_{r}^{3}+2\mathrm{s}_{2}e_{r}^{3}+16\left( \mathrm{c}%
_{1\left( 2\right) }^{2PN}-\mathrm{c}_{2\left( 1\right) }^{2PN}\right) 
\notag \\
&&-16e_{r}\mathrm{c}_{2\left( 2\right) }^{2PN}-16\allowbreak e_{r}^{2}%
\mathrm{c}_{2\left( 1\right) }^{2PN}+16e_{r}\left( \mathrm{c}_{1\left(
1\right) }^{2PN}-\mathrm{c}_{2\left( 0\right) }^{2PN}\right)  \notag \\
&&+2\mathrm{c}_{1\left( 0\right) }^{PN}\left( 6\mathrm{q}_{3}e_{r}+\frac{d%
\mathrm{q}_{1}}{de_{r}}\right) e_{r}-2\mathrm{c}_{2\left( 2\right) }^{PN}%
\frac{d\mathrm{q}_{0}}{de_{r}}  \notag \\
&&-2\mathrm{c}_{2\left( 1\right) }^{PN}\left( \frac{d\mathrm{q}_{1}}{de_{r}}+%
\frac{d\mathrm{q}_{0}}{de_{r}}\right) e_{r}-2\mathrm{c}_{2\left( 0\right)
}^{PN}\frac{d\mathrm{q}_{1}}{de_{r}}e_{r}^{2}  \notag \\
&&+2\left( \mathrm{c}_{1\left( 1\right) }^{PN}-\mathrm{c}_{2\left( 0\right)
}^{PN}\right) \left[ 2\left( 2\mathrm{q}_{2}-\mathrm{q}_{1}\right) e_{r}+%
\frac{d\mathrm{q}_{1}}{de_{r}}e_{r}^{2}+\frac{d\mathrm{q}_{0}}{de_{r}}\right]
\notag \\
&&+2\left( \mathrm{c}_{1\left( 2\right) }^{PN}-\mathrm{c}_{2\left( 1\right)
}^{PN}\right) \left[ 2\left( \mathrm{q}_{1}-2\mathrm{q}_{0}\right) +\frac{d%
\mathrm{q}_{0}}{de_{r}}e_{r}\right] ~,
\end{eqnarray}%
where the terms in the last line cancel by virtue of the relations between
the coefficients $\mathrm{c}_{i(k)}^{PN}$, while the first five lines add up
to zero after inserting the definitions of the coefficients $\mathrm{c}%
_{i(k)}^{PN}$, $\mathrm{c}_{i(k)}^{2PN}$, $\mathrm{q}_{i}$, $\mathrm{s}_{i}$.

The coefficient of $\cos ^{3}\chi _{p}$ gives:%
\begin{eqnarray}
0 &=&4\mathrm{s}_{4}e_{r}^{4}+3\mathrm{s}_{3}e_{r}^{4}+16\left( \mathrm{c}%
_{1\left( 3\right) }^{2PN}-\mathrm{c}_{2\left( 2\right) }^{2PN}\right)  
\notag \\
&&-16e_{r}\mathrm{c}_{2\left( 3\right) }^{2PN}-16\allowbreak e_{r}^{2}%
\mathrm{c}_{2\left( 2\right) }^{2PN}+16e_{r}\left( \mathrm{c}_{1\left(
2\right) }^{2PN}-\mathrm{c}_{2\left( 1\right) }^{2PN}\right)   \notag \\
&&-2\mathrm{c}_{2\left( 2\right) }^{PN}\left( \frac{d\mathrm{q}_{1}}{de_{r}}+%
\frac{d\mathrm{q}_{0}}{de_{r}}\right) e_{r}-2\mathrm{c}_{2\left( 1\right)
}^{PN}\frac{d\mathrm{q}_{1}}{de_{r}}e_{r}^{2}  \notag \\
&&+2\left( \mathrm{c}_{1\left( 1\right) }^{PN}-\mathrm{c}_{2\left( 0\right)
}^{PN}\right) \left( 6\mathrm{q}_{3}e_{r}+\frac{d\mathrm{q}_{1}}{de_{r}}%
\right) e_{r}+4\mathrm{c}_{1\left( 0\right) }^{PN}\mathrm{q}_{3}e_{r}^{3} 
\notag \\
&&+2\left( \mathrm{c}_{1\left( 2\right) }^{PN}-\mathrm{c}_{2\left( 1\right)
}^{PN}\right) \left[ 2\left( 2\mathrm{q}_{2}-\mathrm{q}_{1}\right) e_{r}+%
\frac{d\mathrm{q}_{1}}{de_{r}}e_{r}^{2}+\frac{d\mathrm{q}_{0}}{de_{r}}\right]
\notag \\
&&+2\left( \mathrm{c}_{1\left( 3\right) }^{PN}-\mathrm{c}_{2\left( 2\right)
}^{PN}\right) \left[ 2\left( \mathrm{q}_{1}-2\mathrm{q}_{0}\right) +\frac{d%
\mathrm{q}_{0}}{de_{r}}e_{r}\right] ~,
\end{eqnarray}%
and the terms in the last two lines cancel by virtue of the relations
between the coefficients $\mathrm{c}_{i(k)}^{PN}$, while the first four
lines add up to zero after inserting the definitions of the coefficients $%
\mathrm{c}_{i(k)}^{PN}$, $\mathrm{c}_{i(k)}^{2PN}$, $\mathrm{q}_{i}$, $%
\mathrm{s}_{i}$.

The coefficient of $\cos ^{4}\chi _{p}$ gives:%
\begin{eqnarray}
0 &=&4\mathrm{s}_{4}e_{r}^{5}-16\allowbreak e_{r}^{2}\mathrm{c}_{2\left(
3\right) }^{2PN}+16e_{r}\left( \mathrm{c}_{1\left( 3\right) }^{2PN}-\mathrm{c%
}_{2\left( 2\right) }^{2PN}\right)   \notag \\
&&-2\mathrm{c}_{2\left( 2\right) }^{PN}\frac{d\mathrm{q}_{1}}{de_{r}}%
e_{r}^{2}+4\left( \mathrm{c}_{1\left( 1\right) }^{PN}-\mathrm{c}_{2\left(
0\right) }^{PN}\right) \mathrm{q}_{3}e_{r}^{3}  \notag \\
&&+e_{r}\left( 5\mathrm{s}_{5}e_{r}^{4}-16\mathrm{c}_{2\left( 4\right)
}^{2PN}\right) +16\left( \mathrm{c}_{1\left( 4\right) }^{2PN}-\mathrm{c}%
_{2\left( 3\right) }^{2PN}\right)   \notag \\
&&+2\left( \mathrm{c}_{1\left( 3\right) }^{PN}-\mathrm{c}_{2\left( 2\right)
}^{PN}\right) \left[ 2\left( 2\mathrm{q}_{2}-\mathrm{q}_{1}\right) e_{r}+%
\frac{d\mathrm{q}_{1}}{de_{r}}e_{r}^{2}+\frac{d\mathrm{q}_{0}}{de_{r}}\right]
\notag \\
&&+2\left( \mathrm{c}_{1\left( 2\right) }^{PN}-\mathrm{c}_{2\left( 1\right)
}^{PN}\right) \left( 6\mathrm{q}_{3}e_{r}+\frac{d\mathrm{q}_{1}}{de_{r}}%
\right) e_{r}~,
\end{eqnarray}%
and the terms in the last three lines cancel by virtue of the relations
between the coefficients $\mathrm{c}_{i(k)}^{PN}$, $\mathrm{c}_{i(k)}^{2PN}$%
, $\mathrm{s}_{i}$, while the first two lines add up to zero after inserting
the definitions of the coefficients $\mathrm{c}_{i(k)}^{PN}$, $\mathrm{c}%
_{i(k)}^{2PN}$, $\mathrm{q}_{i}$, $\mathrm{s}_{i}$.

The coefficient of $\cos ^{5}\chi _{p}$ gives:%
\begin{eqnarray}
0 &=&e_{r}^{2}\left( 5\mathrm{s}_{5}e_{r}^{4}-16\allowbreak \mathrm{c}%
_{2\left( 4\right) }^{2PN}\right) +16\left( \mathrm{c}_{1\left( 5\right)
}^{2PN}-\mathrm{c}_{2\left( 4\right) }^{2PN}\right)  \notag \\
&&+16e_{r}\left( \mathrm{c}_{1\left( 4\right) }^{2PN}-\mathrm{c}_{2\left(
3\right) }^{2PN}\right) +4\left( \mathrm{c}_{1\left( 2\right) }^{PN}-\mathrm{%
c}_{2\left( 1\right) }^{PN}\right) \mathrm{q}_{3}e_{r}^{3}  \notag \\
&&+2\left( \mathrm{c}_{1\left( 3\right) }^{PN}-\mathrm{c}_{2\left( 2\right)
}^{PN}\right) \left( 6\mathrm{q}_{3}e_{r}+\frac{d\mathrm{q}_{1}}{de_{r}}%
\right) e_{r}~,
\end{eqnarray}%
where all terms cancel by virtue of the relations between the coefficients $%
\mathrm{c}_{i(k)}^{PN}$, $\mathrm{c}_{i(k)}^{2PN}$, $\mathrm{s}_{i}$.

The coefficient of $\cos ^{6}\chi _{p}$ gives:%
\begin{equation}
0=4\left( \mathrm{c}_{1\left( 5\right) }^{2PN}-\mathrm{c}_{2\left( 4\right)
}^{2PN}\right) +\left( \mathrm{c}_{1\left( 3\right) }^{PN}-\mathrm{c}%
_{2\left( 2\right) }^{PN}\right) \mathrm{q}_{3}e_{r}^{2}~,
\end{equation}%
where all terms cancel by virtue of the relations between the coefficients $%
\mathrm{c}_{i(k)}^{PN}$, $\mathrm{c}_{i(k)}^{2PN}$.

In summary all these 6 equations reduce to identities, confirming the
consistency condition arising from energy conservation.

The consistency condition (\ref{2PNconsistencyJ}), arising from the total
angular momentum conservation takes the explicit form $0=\sum_{k=0}^{5}h_{%
\left( k\right) }\cos \chi _{p}$, with the coefficients

\begin{eqnarray}
h_{\left( 0\right) } &=&\mathrm{b}_{2\left( 0\right) }\mathrm{c}_{2\left(
0\right) }^{PN}-\mathrm{b}_{1\left( 0\right) }\mathrm{c}_{1\left( 0\right)
}^{PN}-2\mathrm{c}_{1\left( 0\right) }^{2PN}-\frac{1}{4}\mathrm{p}_{1}e_{r} 
\notag \\
h_{\left( 1\right) } &=&\mathrm{b}_{2\left( 0\right) }\mathrm{c}_{2\left(
1\right) }^{PN}-\mathrm{b}_{1\left( 0\right) }\mathrm{c}_{1\left( 1\right)
}^{PN}+\mathrm{b}_{2\left( 1\right) }\mathrm{c}_{2\left( 0\right) }^{PN}-%
\mathrm{b}_{1\left( 1\right) }\mathrm{c}_{1\left( 0\right) }^{PN}  \notag \\
&&+2\left( \mathrm{c}_{2\left( 0\right) }^{2PN}-\mathrm{c}_{1\left( 1\right)
}^{2PN}\right) -\left( \frac{1}{2}\mathrm{p}_{2}+\frac{1}{4}\mathrm{p}%
_{1}\right) e_{r}^{2}  \notag \\
h_{\left( 2\right) } &=&\mathrm{b}_{2\left( 0\right) }\mathrm{c}_{2\left(
2\right) }^{PN}-\mathrm{b}_{1\left( 0\right) }\mathrm{c}_{1\left( 2\right)
}^{PN}+\mathrm{b}_{2\left( 1\right) }\mathrm{c}_{2\left( 1\right) }^{PN}-%
\mathrm{b}_{1\left( 1\right) }\mathrm{c}_{1\left( 1\right) }^{PN}  \notag \\
&&+\mathrm{b}_{2\left( 2\right) }\mathrm{c}_{2\left( 0\right) }^{PN}+2\left( 
\mathrm{c}_{2\left( 1\right) }^{2PN}-\mathrm{c}_{1\left( 2\right)
}^{2PN}\right) \!-\!\!\left( \!\frac{3}{4}\mathrm{p}_{3}\!+\!\frac{1}{2}%
\mathrm{p}_{2}\!\right) e_{r}^{3}  \notag \\
h_{\left( 3\right) } &=&\mathrm{b}_{2\left( 1\right) }\mathrm{c}_{2\left(
2\right) }^{PN}-\mathrm{b}_{1\left( 1\right) }\mathrm{c}_{1\left( 2\right)
}^{PN}-\mathrm{b}_{1\left( 0\right) }\mathrm{c}_{1\left( 3\right) }^{PN}+%
\mathrm{b}_{2\left( 2\right) }\mathrm{c}_{2\left( 1\right) }^{PN}  \notag \\
&&+2\left( \mathrm{c}_{2\left( 2\right) }^{2PN}-\mathrm{c}_{1\left( 3\right)
}^{2PN}\right) -\frac{3}{4}\mathrm{p}_{3}e_{r}^{4}  \notag \\
h_{\left( 4\right) } &=&\mathrm{b}_{2\left( 2\right) }\mathrm{c}_{2\left(
2\right) }^{PN}-\mathrm{b}_{1\left( 1\right) }\mathrm{c}_{1\left( 3\right)
}^{PN}+2\left( \mathrm{c}_{2\left( 3\right) }^{2PN}-\mathrm{c}_{1\left(
4\right) }^{2PN}\right)  \notag \\
h_{\left( 5\right) } &=&2\left( \mathrm{c}_{2\left( 4\right) }^{2PN}-\mathrm{%
c}_{1\left( 5\right) }^{2PN}\right) ~,
\end{eqnarray}%
all of which vanishing by virtue of the relations between the coefficients $%
\mathrm{c}_{i(k)}^{PN}$, $\mathrm{c}_{i(k)}^{2PN}$, $\mathrm{b}_{i\left(
k\right) }$ and $\mathrm{p}_{i}$. Therefore the consistency condition
arising from total angular momentum conservation also holds.

\end{document}